%% file: manuscript.tex
\PassOptionsToPackage{hyphens}{url}
\documentclass[sigconf, nonacm, balance=false]{acmart}

\usepackage{textcase}
\makeatletter
\expandafter\def\expandafter\@secfont\expandafter{\@secfont\MakeTextUppercase}
\makeatother

\input{./sections/package.tex}

\newcommand{\sys}{Tessera\xspace}
\newcommand{\spalm}{SPALM\xspace}
\newcommand{\duckdb}{DuckDB\xspace}
\newcommand{\pg}{PostgreSQL\xspace}
\newcommand{\sqlite}{SQLite\xspace}
\newcommand{\spark}{Spark\xspace}
\newcommand{\grb}{GraphBLAS\xspace}

\begin{document}
\title{Tessera: Fast Join-Aggregates over Skewed Sparse Relations in Database Engines}

\author{Gyuyeong Kim}
\affiliation{%
  \institution{Yonsei University}
  \city{Seoul}
  \country{Republic of Korea}
}
\email{gyuyeong.kim@yonsei.ac.kr}

\begin{abstract}
\input{./sections/abstract.tex}
\end{abstract}

\maketitle

\pagestyle{plain}
\begingroup\small\noindent\raggedright\textbf{Artifact Availability:}\\
The source code is available at \url{https://github.com/GyuyeongKim/Tessera-public}.
\endgroup

\input{./sections/introduction.tex}
\input{./sections/motivation.tex}
\input{./sections/design.tex}

\input{./sections/implementation.tex}

\input{./sections/evaluation.tex}
\FloatBarrier
\input{./sections/relatedwork.tex}
\input{./sections/conclusion.tex}
\input{./sections/ack.tex}
\bibliographystyle{ACM-Reference-Format}
\bibliography{./gykim,./tessera}
\end{document}

%% file: sections/package.tex
\input{figures/tikz/tsfigstyle.tex}

\usepackage{xurl}
\hypersetup{breaklinks=true}
\usepackage{epsfig,graphicx,endnotes,subfig,multirow,algorithm,algpseudocode}
\usepackage{xcolor,xspace,booktabs}
\usepackage{pifont}
\usepackage{colortbl}
\usepackage{enumitem}
\usepackage[T1]{fontenc}
\usepackage{color,soul}
\usepackage{threeparttable}
\usepackage{placeins}
\newcommand{\cmark}{\textcolor{green!60!black}{\ding{51}}}
\newcommand{\xmark}{\textcolor{red!80!black}{\ding{55}}}
\DeclareRobustCommand{\pmark}{\textcolor{orange!85!black}{\tikz[baseline=-0.55ex]{\draw[line width=0.5pt] (0,0) circle (0.8ex); \fill (0,0.8ex) arc (90:270:0.8ex) -- cycle;}}}

%% file: figures/tikz/tsfigstyle.tex
\usepackage{tikz}
\usetikzlibrary{positioning,arrows.meta,calc,patterns,decorations.pathreplacing,fit,backgrounds,shadows}
\newcommand{\figfontS}{\fontfamily{Crlt-LF}\fontsize{8}{9.2}\selectfont}
\definecolor{kbRed}{HTML}{D85A4D}
\definecolor{kbRedL}{HTML}{F2BFB7}
\definecolor{kbBlue}{HTML}{3D6FA0}
\definecolor{kbGreen}{HTML}{4A9D7A}
\definecolor{kbGold}{HTML}{D9A65A}
\definecolor{kbViolet}{HTML}{8E72B5}
\definecolor{kbGray}{HTML}{7F7F7F}
\definecolor{tsInk}{HTML}{1F2328}
\definecolor{tsArrow}{HTML}{4A4F57}
\definecolor{tsGray}{HTML}{A3AAB4}
\definecolor{tsGrayL}{HTML}{F3F4F6}
\definecolor{tsGrayM}{HTML}{DDE0E5}
\definecolor{tsSlate}{HTML}{7D8FA8}
\definecolor{tsSlateL}{HTML}{EEF2F7}
\definecolor{tsSlateM}{HTML}{D5DDE8}
\definecolor{tsBlue}{HTML}{2F5E9E}
\definecolor{tsBlueL}{HTML}{E6EDF7}
\definecolor{tsBlueM}{HTML}{B5C8E6}
\definecolor{tsCoral}{HTML}{E07A5F}
\definecolor{tsCoralL}{HTML}{FCEEEA}
\definecolor{tsCoralM}{HTML}{F6CBBE}
\definecolor{tsLav}{HTML}{8C72B8}
\definecolor{tsLavL}{HTML}{F2EEF9}
\definecolor{tsLavM}{HTML}{D9CFEC}
\definecolor{tsRed}{HTML}{B3261E}
\definecolor{tsRedL}{HTML}{F6D5D2}
\tikzset{
  tsbox/.style={draw=#1, fill=#1L, line width=0.6pt, rounded corners=3pt, font=\figfontS, text=tsInk,
                align=center, inner sep=2.2pt, minimum height=5.2mm},
  tsdata/.style={draw=#1, fill=#1L, line width=0.6pt, font=\figfontS, text=tsInk, align=center, inner sep=1.6pt},
  tspill/.style={draw=#1, fill=#1L, line width=0.6pt, rounded corners=4.5pt, font=\figfontS, text=tsInk,
                 align=center, inner sep=1.6pt, minimum height=3.8mm},
  tsregion/.style={draw=#1, fill=#1L, line width=0.6pt, rounded corners=4pt, drop shadow={opacity=0.12,
                   shadow xshift=0.4pt, shadow yshift=-0.5pt}},
  tstitle/.style={font=\figfontS\bfseries, text=tsInk, inner sep=1pt, anchor=north west},
  tslab/.style={font=\figfontS, text=tsInk, inner sep=0.8pt, align=center},
  tsgo/.style={-{Latex[length=1.6mm,width=1.2mm]}, line width=0.7pt, draw=tsArrow},
  tsthin/.style={-{Latex[length=1.4mm,width=1.0mm]}, line width=0.5pt, draw=tsArrow},
  tsstep/.style={circle, fill=tsInk, text=white, inner sep=0pt, minimum size=3.2mm, font=\figfontS\bfseries},
  tscell/.style={draw=#1, fill=#1M, line width=0.4pt},
  tsgrid/.style={draw=tsGray, line width=0.25pt},
  tsbrace/.style={decorate, decoration={brace, amplitude=1.1mm}, draw=tsArrow, line width=0.5pt},
}
\DeclareRobustCommand{\tsn}[1]{\texorpdfstring{\tikz[baseline=(c.base)]\node[tsstep] (c) {#1};}{(#1)}}

%% file: sections/abstract.tex
Join-aggregate queries count or sum matching rows for each pair of values, often over skewed sparse relations in which a few values are frequent and most pairs of values never co-occur. Conventional execution plans materialize billions of intermediate join rows, taking minutes to tens of minutes per query. Recent matrix-based solutions reformulate these queries as matrix multiplication to reduce or avoid this materialization. However, they remain slow because they either store result cells for non-matching pairs or still materialize intermediate join rows for infrequent values. We present \sys, a skew-aware join-aggregate operator for database engines that avoids both sources of overhead. Specifically, \sys computes a sparse matrix product directly over the engine's rows, emits only nonzero cells as result rows, and prunes values that cannot meet the query's threshold before multiplication begins. It also offloads the dense block formed by pairs of the most frequent values to a CPU matrix accelerator when that block is dense enough.
Experimental results show that \sys is faster than the fastest existing system by 2.8$\times$ to 18.8$\times$.
In addition, \sys outperforms DuckDB's execution plans by up to 1,802$\times$.

%% file: sections/introduction.tex
\section{Introduction \label{sec:intro}}
Many analytics questions ask how often two things occur together.
For example, an analyst asks which words appear in the same documents, which movies the same users rate, which products the same customers buy, or how many friends two users share.
In SQL, this is a \emph{join-aggregate} query.
It joins a relation with itself, or with a second relation, on a shared key, and then groups the joined rows by one attribute from each side to count or sum them.
Many of these queries retain only strong pairs whose counts meet a \emph{threshold}, such as words that co-occur in at least 100 documents, or rank among the top $k$.
Join-aggregates appear in decision-support benchmarks~\cite{tpcxbb,ldbcbi} and are used to build set similarity joins and item-to-item recommendation~\cite{ssjbench,sarplus}.
In these use cases, the execution time of a join-aggregate matters.
This is because an analyst waits for each answer before asking the next question, and a recommender recomputes its item similarities every time it retrains on new data.

The relations behind these questions are skewed and sparse.
Value frequencies are heavy-tailed~\cite{clauset}, with a few values, such as common words or popular movies, in many rows and most values in few.
Most pairs of values also never co-occur, since no key holds both of them.
We call them \emph{non-matching pairs}.
For example, in a collection of 300K news articles~\cite{uciBoW}, 89\% of the pairs of words never appear in the same article.
However, current database engines perform join-aggregates slowly on such relations.
For example, counting word co-occurrence over the news articles produces 20 billion joined rows, and an analytical engine such as \duckdb~\cite{duckdb} builds and aggregates every one of them.

Prior work reduces or avoids intermediate join rows by computing a join-aggregate as \emph{a matrix product} and accelerates queries by speeding up multiplication~\cite{deepjoinproject,humatmul,spalm,tcudb}. However, these designs remain inefficient on skewed sparse relations.
Specifically, \emph{join-project algorithms}~\cite{deepjoinproject,humatmul} multiply frequent values as a dense matrix but still materialize join rows for rare values.
Meanwhile, \emph{in-database linear algebra}~\cite{spalm} keeps dense results with cells for non-matching pairs, and \emph{tensor-core databases}~\cite{tcudb} either do the same or switch the whole product to a sparse kernel.
These costs limit the benefit of faster multiplication, even with recent on-CPU matrix accelerators (e.g., Intel AMX~\cite{intelamx}).
In this context, we ask: \emph{how can a database engine speed up the whole join-aggregate on skewed sparse relations?}

To answer this question affirmatively, we present \sys, a high-performance skew-aware join-aggregate operator for database engines.
Our key insight is that the frequent values and the rare values cost differently.
Since frequent values appear together in many keys, each of their pairs is counted many times, and their pairs receive a large share of the products.
We call these few values the \emph{dense core}.
In contrast, the rare values, which form the \emph{long tail} of the frequency distribution, are far more numerous, and they produce most of the result cells, each with few products.
Therefore, accelerating the core pays off only when processing the long tail and writing the results are also cheap.
Based on this insight, \sys splits the values at a \emph{dense frontier}, which separates the dense core of frequent values from the long tail.
Beyond the frontier, a sparse path computes only the nonzero result cells of rarer values and does most of the work.
Inside the frontier, a matrix accelerator multiplies the pairs of frequent values as dense tiles.
\sys chooses the frontier for each query by comparing the cost of each accelerator tile with that of the sparse products it replaces.

Realizing this design raises four technical challenges.
First, the sparse part must be fast on its own, since the rare values and the output carry most of the work.
The output can also have more rows than the engine's memory budget holds.
Furthermore, a query that keeps only the pairs whose counts meet a threshold usually does as much work as one that keeps all pairs, since the threshold applies only after each cell is computed.
Second, the best frontier depends on the relation and on the query, because the accelerator pays for whole tiles and pruning changes which values remain.
Third, the matrix accelerator multiplies dense tiles, but the input is a set of rows.
Turning the frequent values into a dense matrix with one row per key wastes the accelerator, since each key holds only a few of them and most entries are zeros.
This matrix also grows with the number of keys rather than with the result.
Lastly, the operator must run inside engines that differ in how they plan queries and manage resources.

\sys addresses these challenges with the following mechanisms.
First, the sparse path builds on Gustavson's sparse product algorithm~\cite{gustavson} but computes only what a relational result needs (\S\ref{sec:design-sparse}).
It adds products into counters instead of building joined rows, computes each pair of values once, and appends the nonzero cells as unordered rows instead of assembling a sorted matrix.
These rows spill to files when they exceed the engine's memory limit.
Before any product, it drops the values that cannot reach the threshold, and a stricter query therefore runs faster.
Second, a cost model scans the rows once, and \sys picks the frontier with the lowest predicted runtime after pruning (\S\ref{sec:design-frontier}).
If no block is dense enough, the accelerator costs only this scan.
Third, \sys groups rows that share frequent values and skips tiles without useful products (\S\ref{sec:design-grid}).
As a result, the accelerator's work follows its nonzero tiles rather than the size of the core.
Lastly, each engine reaches \sys through a C interface and a small adapter built on its extension hooks for plans, scans, and typed output (\S\ref{sec:design-engines}).

\sys is readily deployable.
We integrate \sys into four database engines as extensions, \duckdb, \pg~\cite{postgresql}, \sqlite~\cite{sqlite}, and \spark~\cite{sparksql}.
In \duckdb, \pg, and \spark, the adapter rewrites the plan of a join-aggregate, whose query therefore needs no change.
Users can install \sys through each engine's own package channel.
We conduct a series of experiments and find that \sys outperforms the fastest existing system by 2.8$\times$ to 18.8$\times$.
In all four engines, \sys is faster than their own plans, especially by up to 1,802$\times$ in \duckdb.

In summary, this work makes the following contributions:
\begin{itemize}[noitemsep,leftmargin=*]
\item We show that on skewed sparse relations, a faster multiplication, even on a matrix accelerator, gains little, since existing matrix-based designs still spend their work on non-matching pairs and on the rows of rare values.
\item We design \sys, a skew-aware operator that runs a sparse product on the engine's rows and returns rows, prunes values before the product, and gives a matrix accelerator the dense core that a cost model sizes for each query.
\item We implement \sys in four database engines and show that it outperforms both existing matrix-based systems and the engines' own plans.
\end{itemize}

%% file: sections/motivation.tex
\section{Background and Motivation \label{sec:motiv}}

\subsection{Join-Aggregates \label{sec:motiv-ja}}

\noindent\textbf{The query.}
A join-aggregate over a relation $R(k, x)$ joins $R$ with itself on the key $k$ and counts the joined rows for each pair of values of $x$:
\begin{center}\small
\texttt{SELECT a.x, b.x, COUNT(*) FROM R a, R b}\\
\texttt{WHERE a.k = b.k GROUP BY a.x, b.x}
\end{center}
In a word co-occurrence query, $k$ is a document and $x$ is a word.
In item-to-item recommendation, $k$ is a user and $x$ is an item, and in common-neighbor counting, $k$ and $x$ are both vertices of a graph.

Variants of the query sum a product of two weights, \texttt{SUM(a.v * b.v)}, return one row per unordered pair with \texttt{a.x < b.x}, or select only the strong pairs with \texttt{HAVING} or \texttt{ORDER BY ... LIMIT}.
Frequent pair mining is the last form with a support threshold~\cite{apriori,fimi}.
A \emph{two-relation product} joins $R(k, x)$ with a second relation $S(k, y)$ and groups by $x$ and $y$, such as movies and the tags that their viewers use.
Item similarity divides the count of a pair by the sizes of the two items, as in cosine or Jaccard similarity, and reports the pairs above a threshold~\cite{allpairs,ssjbench}.

\begin{figure}[t]
\centering
\subfloat[Hash join and aggregation, one hash-table update per joined row]{\input{figures/tikz/fig_joinrows.tex}}\\[4pt]
\subfloat[Matrix product, no joined rows but a dense result]{\input{figures/tikz/fig_matrix.tex}}
\caption{Two ways to compute a join-aggregate, which waste work on (a) joined rows or (b) zero result cells.\label{fig:matrix}}
\vspace{-6mm}
\end{figure}
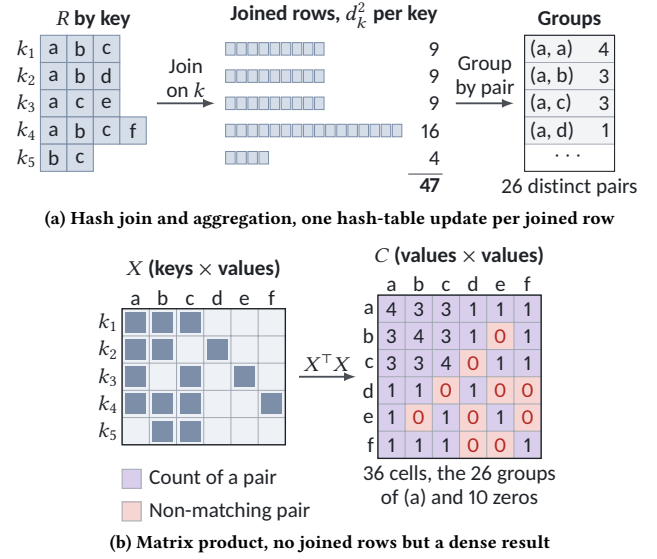

\noindent\textbf{Why it is expensive.}
A database engine runs the query as a hash join followed by a hash aggregation, making every joined row a hash-table update (Figure~\ref{fig:matrix}~(a)).
The join emits $d_k^2$ rows for a key that occurs in $d_k$ rows.
The engine processes $\sum_k d_k^2$ joined rows, a sum that the few long keys of skewed data dominate.
For example, the 300K articles of a news collection hold 70 million distinct document-word pairs but produce 20 billion joined rows, almost all of them from long documents and common words.
Thresholds do not help, since the engine filters the groups only after it aggregates every joined row.
Pushing the aggregation below the join~\cite{eageragg} does not remove these rows either, since each side must keep both its key and its value and can merge only repeated rows.

\noindent\textbf{The matrix view.}
Computing a join-aggregate as a matrix product removes the joined rows, but on real relations most cells of the result are zero.
Let $X$ be the matrix with one row per key and one column per value, where $X_{k,x}$ is the number of rows $(k, x)$, or the weight $v$.
The count of the pair $(i, j)$ is then $C_{i,j} = \sum_k X_{k,i} X_{k,j}$.
Hence, $C = X^\top X$, and a two-relation product is $X^\top Y$.
As Figure~\ref{fig:matrix}~(b) shows, the product reads each row of $X$ once and adds its products into $C$ instead of building joined rows.
However, both $X$ and $C$ are sparse.
Existing matrix-based designs nonetheless focus on how fast this product runs.

\subsection{Matrix-Based Designs \label{sec:motiv-limits}}\label{sec:motiv-case}

\noindent\textbf{Existing matrix-based designs.}
Designs that compute a join-aggregate as a matrix product differ mainly in how they multiply, and each leaves the rare values or the result on a slower path.
Join-project algorithms split the values by frequency and run the frequent part as a dense matrix product~\cite{humatmul,deepjoinproject}.
However, they still join the rare values and build their joined rows.
In-database linear algebra such as \spalm~\cite{spalm} picks a sparse or a dense kernel for each block inside \duckdb~\cite{duckdb}, and its \duckdb operator writes a dense result of all pairs.
Tensor-core databases such as TCUDB~\cite{tcudb} run dense products on GPUs and fall back to a hash join or a sparse kernel on sparse inputs.
Sparse matrix libraries such as \grb~\cite{graphblas}, which some engines embed~\cite{redisgraph}, handle sparse operands well.
However, they work on matrices in their own format and assemble the result as a matrix.
\spalm, TCUDB, and \grb also filter by a threshold only after the full product.

\begin{figure}[t]
\centering
\includegraphics[scale=0.5]{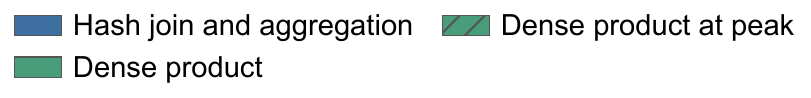}\\
\subfloat[News articles]{\includegraphics[width=0.48\linewidth]{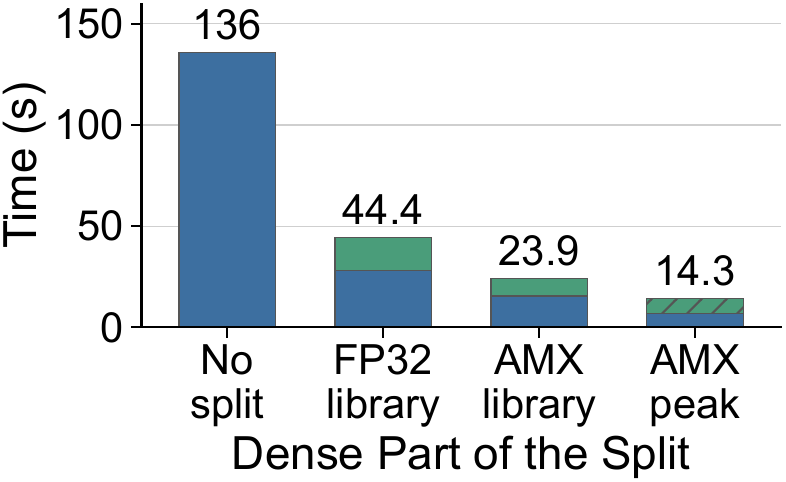}}\hfill
\subfloat[Medical abstracts]{\includegraphics[width=0.48\linewidth]{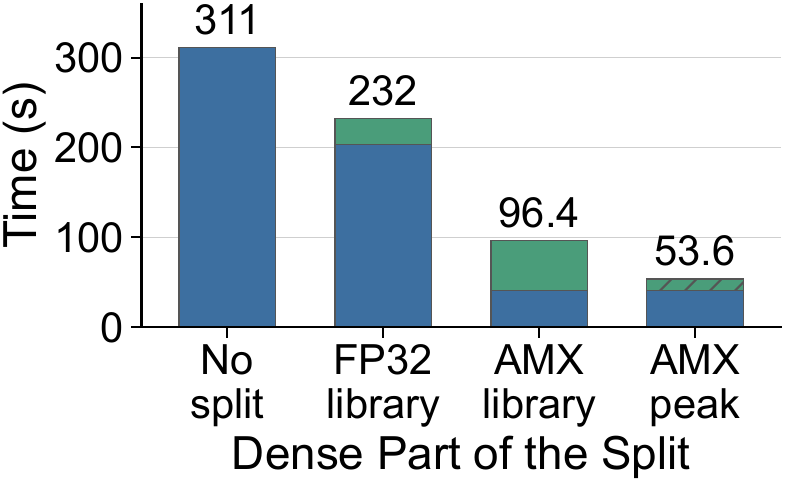}}
\caption{Offloading only the frequent values to AMX, even at its peak speed, leaves much of the time to the rare values. \textit{The rest runs in \duckdb~1.5.6 with \texttt{a.x < b.x}, and AMX peak charges the dense part at the peak rate of tile products whose operands stay in the tile registers.}
\label{fig:natural}}
\vspace{-3mm}
\end{figure}

\noindent\textbf{Faster multiplication gains little.}
A faster multiplication speeds up matrix-based designs only a little, because the rare values and the result take a large share of their time.
For example, on the block of the 8,192 most frequent news words, \spalm spends only 0.67 of its 2.75 seconds of kernel time in kernels with a dense operand.
Therefore, faster dense kernels speed up the block by at most 1.32$\times$.
MMJoin~\cite{mmjoin}, a standalone implementation of the frequency split, spends only 2.5--16\% of its time in its dense product at eight of nine support thresholds.
We also measure a natural design that runs the frequent block of this split on Intel AMX~\cite{intelamx}, which multiplies small dense matrices in a single CPU instruction.
It uses a vendor library for this block and leaves the rest to the engine.
Figure~\ref{fig:natural} shows that with \texttt{a.x < b.x}, the rest still takes 47\% and 76\% of the time at the best split even if the dense part ran at the peak speed of AMX.
As a result, a matrix accelerator pays off only when the rare values and the result are also cheap.

\subsection{Our Approach \label{sec:motiv-approach}}
\begin{table}[t]
\centering\footnotesize
\caption{Comparison to existing systems (\pmark: in some plans).
\label{tab:comparison}}
\setlength{\tabcolsep}{1.3pt}
\begin{tabular}{@{}llcccc@{}}
\toprule
 & \shortstack[l]{Dense\\unit} & \shortstack{No joined\\rows} & \shortstack{No zero\\result cells} & \shortstack{Threshold\\saves work} & \shortstack{Runs in\\engine} \\
\midrule
\duckdb, \pg~\cite{duckdb,postgresql} & None & \xmark & \cmark & \xmark & \cmark \\
MMJoin~\cite{deepjoinproject,mmjoin} & Vector & \xmark & \cmark & \cmark & \xmark \\
\spalm~\cite{spalm} & Vector & \cmark & \xmark & \xmark & \cmark \\
TCUDB~\cite{tcudb} & GPU & \cmark & \pmark & \xmark & \cmark \\
\grb~\cite{graphblas,redisgraph} & None & \cmark & \cmark & \xmark & \cmark \\
\rowcolor{tsBlueL}\sys & Matrix acc. & \cmark & \cmark & \cmark & \cmark \\
\bottomrule
\end{tabular}
\vspace{-3mm}
\end{table}

\sys follows the skew of the data.
A sparse path handles the rare values and the output, and a CPU matrix accelerator handles the dense core.
The two meet at a frontier that a cost model sets for each query.
Table~\ref{tab:comparison} shows that every existing system misses at least one of these properties, and none of them runs a CPU matrix accelerator.

\noindent\textbf{The sparse path.}
The sparse path is the main path of \sys, since the rare values produce most of the cells of the result.
Any design must compute their products and write their cells, and \sys makes each of these steps cheap.
For example, to count the words that appear with one word, \sys walks the articles that contain it and adds to a counter for each other word in them.
It builds no joined row and probes no hash table.
Only the counters that a value touched become result rows, which go unsorted into per-thread chunks.
A sparse matrix library computes the same counts~\cite{gustavson,graphblas}, but \sys emits the cells as rows instead of a matrix.
Hence, its cost follows the products and the nonzero cells, never all pairs of values or a dense result.

\noindent\textbf{Pruning before the product.}
Two words cannot share more articles than either of them appears in, and with weights or repeated rows, this bound is multiplied by the largest entry of $X$.
Therefore, with a threshold, \sys drops the values that cannot reach it before any product.
For a top-$k$ query, a first pass over the most frequent values yields such a threshold.
As a result, a stricter query runs faster.

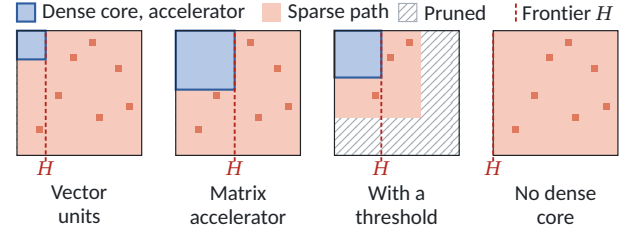
\begin{figure}[t]
\centering
\input{figures/tikz/fig_frontier_moves.tex}
\caption{\sys determines the dense frontier $H$ per query and gives the accelerator only the frequent values. \label{fig:frontiermove}}
\vspace{-3mm}
\end{figure}

\noindent\textbf{The dense core on a matrix accelerator.}
A \emph{matrix accelerator}, such as Intel AMX~\cite{intelamx}, Arm SME~\cite{armsme}, or the matrix facility of IBM Power10~\cite{power10mma}, is a unit of a CPU that multiplies small matrices, called \emph{tiles}, in one instruction.
It works on the CPU's own caches and memory and multiplies several times faster than the vector units.
However, it always multiplies whole tiles, and a zero in a tile costs as much as a useful product.
Therefore, a block of the result pays off on the accelerator only once its \emph{density}, its share of useful products, exceeds a break-even density $d^*$.
$d^*$ is the accelerator's time per multiply-add divided by the time of one sparse product.
Skew forms blocks above this density once the values are ordered by frequency.
For example, the block of the 4,096 most frequent news words holds 48\% of all joined rows and lies above $d^*$, unlike that of the medical abstracts.
Hence, whether a core pays off depends on the relation.

\noindent\textbf{Where the two parts meet.}
How \sys divides the pairs of values between the two parts decides how much the accelerator saves.
Since the blocks above $d^*$ gather among the most frequent values, \sys divides them at one point of the frequency order, the \emph{dense frontier}, which is the number of most frequent values whose pairs run on the accelerator.
A frontier that is too small leaves dense blocks to the sparse path, and one that is too large gives the accelerator tiles that hold mostly zeros.
Moreover, the best frontier moves, as Figure~\ref{fig:frontiermove} shows.
Since a faster accelerator lowers $d^*$ and a cheaper sparse path raises it, the first moves the frontier outward and the second moves it inward.
Pruning moves it too, since it removes values first.
Therefore, a cost model places the frontier for each query where a tile costs as much as the sparse products it replaces.

\noindent\textbf{Why it works.}
The sparse path builds no joined rows, but it still performs one product for each pair of values in a key, about half as many products as the engine has joined rows.
Its gain is therefore a constant factor per product, since an addition into a counter replaces a joined row and a hash-table update, and it writes only the nonzero cells.
The accelerator runs only where its tiles cost less than the sparse products they replace.
When no block is dense enough, \sys runs only the sparse path, after the model's pass over the rows.
Pruning drops a value only when none of its cells can reach the threshold, which keeps the result exact.

%% file: figures/tikz/fig_joinrows.tex
\begin{tikzpicture}[x=0.35cm, y=0.36cm, font=\figfontS]
  \foreach \k/\vals/\n [count=\r from 0] in {1/{a,b,c}/9, 2/{a,b,d}/9, 3/{a,c,e}/9, 4/{a,b,c,f}/16, 5/{b,c}/4} {
    \node[tslab, anchor=east] at (-0.05,4.5-\r) {$k_{\k}$};
    \foreach \v [count=\c from 0] in \vals {
      \draw[tscell=tsSlate] (\c,4-\r) rectangle (\c+1,5-\r);
      \node[tslab] at (\c+0.5,4.5-\r) {\v};
    }
    \def\pos{0}
    \foreach \u in \vals {
      \foreach \w in \vals {
        \pgfmathsetmacro{\px}{7.0+0.42*\pos}
        \fill[tsSlateM] (\px,4.27-\r) rectangle (\px+0.36,4.73-\r);
        \draw[tsSlate, line width=0.3pt] (\px,4.27-\r) rectangle (\px+0.36,4.73-\r);
        \pgfmathsetmacro{\npos}{\pos+1}
        \xdef\pos{\npos}
      }
    }
    \node[tslab, anchor=east] at (15.2,4.5-\r) {\n};
  }
  \node[tstitle, anchor=south] at (2,5.15) {$R$ by key};
  \node[tstitle, anchor=south] at (11.1,5.15) {Joined rows, $d_k^2$ per key};
  \draw[tsArrow, line width=0.4pt] (13.9,-0.1) -- (15.2,-0.1);
  \node[tslab, anchor=north east, font=\figfontS\bfseries] at (15.2,-0.15) {47};
  \draw[tsgo] (4.4,2.5) -- node[tslab, above=1pt] {Join\\on $k$} (6.6,2.5);
  \draw[tsgo] (15.7,2.5) -- node[tslab, above=1pt] {Group\\by pair} (17.9,2.5);
  \foreach \g/\n [count=\r from 0] in {{(a, a)}/4, {(a, b)}/3, {(a, c)}/3, {(a, d)}/1} {
    \draw[tsgrid, fill=tsGrayL] (18.3,4-\r) rectangle (21.7,5-\r);
    \node[tslab, anchor=west] at (18.4,4.5-\r) {\g};
    \node[tslab, anchor=east] at (21.6,4.5-\r) {\n};
  }
  \node[tslab] at (20.0,0.5) {$\cdots$};
  \draw[tsInk, line width=0.4pt] (18.3,0) rectangle (21.7,5);
  \node[tstitle, anchor=south] at (20.0,5.15) {Groups};
  \node[tslab, anchor=north] at (20.0,-0.15) {26 distinct pairs};
\end{tikzpicture}

%% file: figures/tikz/fig_matrix.tex
\begin{tikzpicture}[x=0.36cm, y=0.36cm, font=\figfontS]
  \fill[tsSlateL] (0,0) rectangle (6,5);
  \foreach \c/\r in {0/4,1/4,2/4, 0/3,1/3,3/3, 0/2,2/2,4/2, 0/1,1/1,2/1,5/1, 1/0,2/0}
    \fill[tsSlate] (\c+0.12,\r+0.12) rectangle (\c+0.88,\r+0.88);
  \draw[tsgrid] (0,0) grid[step=1] (6,5);
  \draw[tsInk, line width=0.5pt] (0,0) rectangle (6,5);
  \foreach \v [count=\i from 0] in {a,b,c,d,e,f} \node[tslab, anchor=south] at (\i+0.5,5.05) {\v};
  \foreach \k [count=\i from 0] in {5,4,3,2,1} \node[tslab, anchor=east] at (-0.05,\i+0.5) {$k_{\k}$};
  \node[tstitle, anchor=south] at (3,6.0) {$X$ (keys $\times$ values)};
  \draw[tsgo] (6.5,2.5) -- node[tslab, above=1pt] {$X^\top X$} (8.6,2.5);
  \begin{scope}[shift={(9.4,-0.5)}]
    \fill[tsLavM] (0,0) rectangle (6,6);
    \foreach \c/\r in {1/1,4/4, 2/2,3/3, 3/1,4/2, 3/0,5/2, 4/0,5/1} \fill[tsRedL] (\c,\r) rectangle (\c+1,\r+1);
    \draw[tsgrid] (0,0) grid[step=1] (6,6);
    \draw[tsInk, line width=0.5pt] (0,0) rectangle (6,6);
    \foreach \row [count=\r from 0] in {{4,3,3,1,1,1}, {3,4,3,1,0,1}, {3,3,4,0,1,1}, {1,1,0,1,0,0}, {1,0,1,0,1,0},
                                         {1,1,1,0,0,1}} {
      \foreach \n [count=\c from 0] in \row {
        \ifnum\n=0 \node[tslab, text=tsRed] at (\c+0.5,5.5-\r) {0};
        \else \node[tslab] at (\c+0.5,5.5-\r) {\n}; \fi
      }
    }
    \foreach \v [count=\i from 0] in {a,b,c,d,e,f} {
      \node[tslab, anchor=south] at (\i+0.5,6.05) {\v};
      \node[tslab, anchor=east] at (-0.05,5.5-\i) {\v};
    }
    \node[tstitle, anchor=south] at (3,7.0) {$C$ (values $\times$ values)};
    \node[tslab, anchor=north, align=center] at (3,-0.15) {36 cells, the 26 groups\\of (a) and 10 zeros};
  \end{scope}
  \fill[tsLavM] (0,-1.6) rectangle (0.7,-0.9); \draw[tsgrid] (0,-1.6) rectangle (0.7,-0.9);
  \node[tslab, anchor=west] at (0.9,-1.25) {Count of a pair};
  \fill[tsRedL] (0,-2.8) rectangle (0.7,-2.1); \draw[tsgrid] (0,-2.8) rectangle (0.7,-2.1);
  \node[tslab, anchor=west] at (0.9,-2.45) {Non-matching pair};
\end{tikzpicture}

%% file: figures/tikz/fig_frontier_moves.tex
\begin{tikzpicture}[x=1cm, y=1cm, font=\figfontS]
  \def\s{1.65}
  \foreach \ox/\h/\p/\name [count=\n] in {0/0.38/1.65/{Vector\\units}, 2.1/0.78/1.65/{Matrix\\accelerator},
                                          4.2/0.62/1.15/{With a\\threshold}, 6.3/0/1.65/{No dense\\core}} {
    \begin{scope}[shift={(\ox,0)}]
      \fill[pattern=north east lines, pattern color=tsGray] (0,0) rectangle (\s,\s);
      \fill[tsCoralM] (0,\s-\p) rectangle (\p,\s);
      \ifdim\h pt>0pt
        \fill[tsBlueM] (0,\s-\h) rectangle (\h,\s);
        \draw[tsBlue, line width=0.8pt] (0,\s-\h) rectangle (\h,\s);
      \fi
      \foreach \x/\y in {0.95/1.45, 1.3/1.1, 0.5/0.75, 1.05/0.45, 0.25/0.3, 1.45/0.6, 0.7/1.25} {
        \pgfmathtruncatemacro{\inside}{(\x<\p && \y>\s-\p && !(\x<\h && \y>\s-\h)) ? 1 : 0}
        \ifnum\inside=1 \fill[tsCoral] (\x,\y) rectangle ++(0.09,0.09);\fi
      }
      \draw[tsInk, line width=0.45pt] (0,0) rectangle (\s,\s);
      \draw[tsRed, line width=0.7pt, dash pattern=on 1.6pt off 1.0pt] (\h,-0.08) -- (\h,\s);
      \node[tslab, text=tsRed, anchor=north, inner sep=0.3pt] at (\h,-0.06) {$H$};
      \node[tslab, anchor=north, align=center] at (\s/2,-0.36) {\name};
    \end{scope}
  }
  \begin{scope}[shift={(0,\s+0.12)}]
    \fill[tsBlueM] (0,0) rectangle (0.24,0.24); \draw[tsBlue, line width=0.8pt] (0,0) rectangle (0.24,0.24);
    \node[tslab, anchor=west] at (0.3,0.12) {Dense core, accelerator};
    \fill[tsCoralM] (3.25,0) rectangle (3.49,0.24);
    \node[tslab, anchor=west] at (3.55,0.12) {Sparse path};
    \fill[pattern=north east lines, pattern color=tsGray] (5.05,0) rectangle (5.29,0.24); \draw[tsGray, line width=0.4pt] (5.05,0) rectangle (5.29,0.24);
    \node[tslab, anchor=west] at (5.35,0.12) {Pruned};
    \draw[tsRed, line width=0.7pt, dash pattern=on 1.6pt off 1.0pt] (6.6,0) -- (6.6,0.24);
    \node[tslab, anchor=west] at (6.65,0.12) {Frontier $H$};
  \end{scope}
\end{tikzpicture}

%% file: sections/design.tex
\section{Design \label{sec:design}}

\subsection{Overview \label{sec:design-overview}}

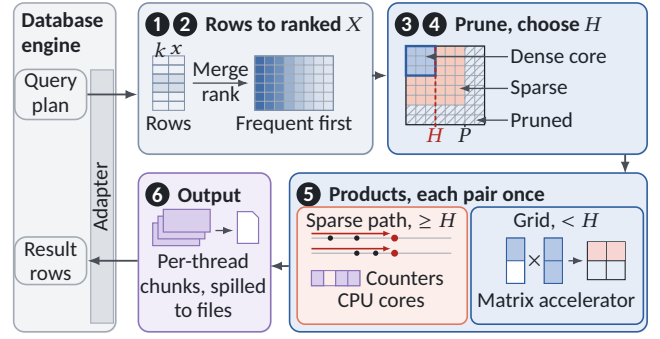
\begin{figure}[t]
\centering
\input{figures/tikz/fig_overview.tex}
\caption{Workflow of \sys. \textit{\tsn{1} The adapter hands the input rows to \sys. \tsn{2} \sys ranks the values by frequency and stores each key's values in rank order. \tsn{3} A threshold removes the values that cannot reach it. \tsn{4} A cost model sets the frontier. \tsn{5} The matrix accelerator and the sparse path compute pairs. \tsn{6} Result rows are returned.}\label{fig:overview}}
\vspace{-3mm}
\end{figure}

\sys is a single operator that a database engine calls in place of its join and aggregation.
It takes the rows $(k, x)$ of the input, or $(k, x, v)$ with a weight, and returns the nonzero cells of $C = X^\top X$ as rows of the result.
Figure~\ref{fig:overview} shows its workflow.

\noindent\tsn{1}~\textbf{Collection.}
The engine's adapter recognizes the join-aggregate in the query plan and streams the input rows to \sys once.
\sys codes the keys as dense integers and merges repeated rows $(k, x)$ into one entry whose value is their count, or the sum of their weights.

\noindent\tsn{2}~\textbf{Ranking.}
\sys numbers the values $x$ by frequency, the most frequent first, and stores $X$ by key, with each row's values in rank order.
After this step, the block of the first $H$ ranks is the densest block of $C$.
Hence, one number $H$ splits the work, and pruning becomes a cut of the rank order.

\noindent\tsn{3}~\textbf{Pruning.}
For a query with \texttt{HAVING} or \texttt{ORDER BY ... LIMIT}, \sys drops the values whose cells cannot pass the threshold.

\noindent\tsn{4}~\textbf{Frontier.}
A cost model picks the frontier $H$ for this query.

\noindent\tsn{5}~\textbf{Products.}
The \emph{grid} computes the pairs of values whose ranks are both below $H$ on the matrix accelerator.
The \emph{sparse path} computes the pairs of two rarer values.
The pairs of a frontier value and a rarer value run on the sparse path for sums and, for counts, on the matrix accelerator through the \emph{heavy-light path}.
Both compute each unordered pair once.

\noindent\tsn{6}~\textbf{Output.}
The cells $(i, j, C_{i,j})$ go to per-thread output chunks.
The adapter maps the ranks back to the original values and emits the rows in the engine's types, with both orders of each pair or one order for \texttt{a.x < b.x}.

\noindent\textbf{Supported queries.}
\sys runs three forms of join-aggregate.
The first is the self-join aggregate \texttt{SELECT a.x, b.x, f FROM R a JOIN R b ON a.k = b.k GROUP BY a.x, b.x}, where $f$ is \texttt{COUNT(*)} or \texttt{SUM(a.v * b.v)}.
It may add a pair filter \texttt{a.x < b.x} (or $\le$, $>$, $\ge$), \texttt{HAVING f >= c}, or \texttt{ORDER BY f DESC LIMIT k}.
The second is \texttt{COUNT(*)} or \texttt{SUM(a.v * b.w)} over two relations $R(k, x, v)$ and $S(k, y, w)$ joined on $k$ and grouped by \texttt{a.x} and \texttt{b.y}.
The third is item similarity above a threshold, with the cosine, Jaccard, Dice, or overlap measure, which \sys offers as a function in every engine.
In all three, $R$ and $S$ are tables or views of one table, keys and values are 64-bit integers or strings without NULLs, and weights are numbers.
For any other query, the adapter leaves the plan unchanged, and the engine runs it with its own join and aggregation.
Examples are \texttt{a.x <> b.x}, a third relation, a further grouping column or aggregate, and row-level security.

\subsection{The Sparse Path and the Output \label{sec:design-sparse}}

The sparse path must compute every pair that involves a rare value and write every nonzero cell of the result.
On skewed relations, these pairs and the output, which can hold billions of cells, take most of the time.
Therefore, the cost of each product and each cell decides the time of the query.
A straightforward approach would be to hand $X$ to a sparse matrix library and convert the resulting matrix back into rows.
However, the library does more work than the query needs.
Specifically, it builds $C$ as a sorted matrix and often computes both halves of the symmetric product, then adds the products via hash tables.
It also applies a threshold only after computing every cell.
The principle of \sys is to compute only what a relational result needs.
A relational result requires only one half of the product and accepts its cells in unordered rows.
Moreover, only values that meet the threshold contribute to it.
Hence, the sparse path scans the rank-ordered rows once per column, removes values before any product, and appends its cells to per-thread chunks.

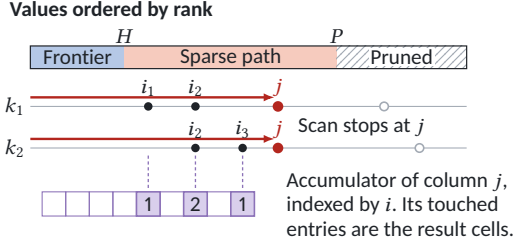
\begin{figure}[t]
\centering
\input{figures/tikz/fig_sparse.tex}
\caption{Example of the symmetric sparse path and item pruning. \textit{Each row is scanned up to the column value $j$, and values beyond the pruning prefix never enter the product.}\label{fig:sparse}}
\vspace{-5mm}
\end{figure}

\noindent\textbf{Symmetric sparse path.}
The sparse path computes one column $j \ge H$ of $C$ at a time with Gustavson's algorithm, and the threads take columns in parallel.
For each row that holds $j$, it adds $X_{k,i} X_{k,j}$ for the row's values $i < j$ into an array indexed by the rank $i$.
In the word example, each article that contains word $j$ adds to the counters of its more frequent words.
Since each row is in rank order, the scan stops at $j$, and each unordered pair is computed once.
For counts, the scan also skips the frontier values, whose pairs with $j$ run on the heavy-light path.
After the last row, the touched entries are the cells of column $j$, and \sys emits them as rows and resets only those entries.

\noindent\textbf{Item pruning.}
Instead of filtering the cells after the product, \sys bounds each value.
For counts and integer weights, $|C_{i,j}|$ is at most the total of $|X|$ in value $i$'s column, the frequency of $i$ for counts, times the largest entry of $|X|$.
\sys takes the maximum of these bounds over each suffix of the ranks.
As a result, the values that can reach the threshold form a prefix.
Since the bounds mostly fall with the frequency rank, a high threshold leaves a short prefix.
Cutting the rest removes values before any product and before the frontier is chosen.
As a result, a stricter query multiplies fewer values and runs faster.
The engine still applies the threshold to the remaining cells above the operator, and the result stays exact.
For \texttt{ORDER BY ... LIMIT k}, the threshold is the $k$-th largest cell, which is unknown before the product.
Since the bounds are largest among the most frequent values, \sys first computes only the cells among them and takes the $k$-th largest of these cells.
The true $k$-th largest cell cannot be smaller, and this value therefore serves as the threshold.
A second run then adds only the cells of the other values that can reach it.

\noindent\textbf{Output chunks.}
The threads must write many cells quickly, and the output may not fit in the memory that the engine allows.
A single shared buffer would make the threads wait for each other.
Separate per-thread results merged at the end would copy every cell again.
Instead, each thread writes its cells into its own fixed-size blocks, called \emph{chunks}, and \sys hands these chunks to the engine as the result without merging them.
When the output exceeds the engine's memory limit, each thread moves every full chunk to its own temporary file and reuses the chunk's memory.
Hence, the output's memory stays close to the limit however large the output grows.
The engine reads the moved chunks back from the files, as engines do when a large sort or hash table spills.

\noindent\textbf{An example of the sparse path.}
In Figure~\ref{fig:sparse}, value $j$ occurs in rows $k_1$ and $k_2$.
Row $k_1$ adds its products with $i_1$ and $i_2$, and row $k_2$ with $i_2$ and $i_3$.
As a result, column $j$ ends with three cells, one of them with a count of 2.

\subsection{Choosing the Dense Frontier \label{sec:design-frontier}}

\sys must decide for each query how many of the most frequent values run on the matrix accelerator.
The choice is hard for three reasons.
First, the accelerator pays per tile, and how many tiles hold a nonzero depends on how the rows are grouped, not only on the density of a block.
Second, the frontier carries a setup cost, since \sys must order the rows that it covers and build their tiles.
Third, the same table needs different frontiers for different thresholds, because pruning changes the values that remain.
Therefore, a fixed rule, such as a number of values or a density threshold per block, would suit some relations and queries and miss on others.
We address this challenge by estimating the time of every candidate frontier from the rows themselves, cheaply enough to do so before each query.
Specifically, a cost model evaluates all candidates in one pass over the rows and picks the cheapest.

\begin{figure}[t]
\centering
\input{figures/tikz/fig_frontier.tex}
\caption{Example of the estimated time of candidate frontiers for a text collection. \textit{Above 1K values, the tile term grows slowly while the sparse term falls, until the frontier reaches rare values.}\label{fig:frontier}}
\vspace{-3mm}
\end{figure}
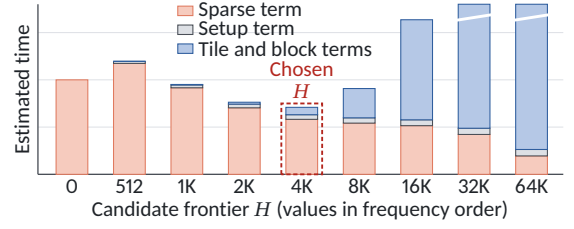

\noindent\textbf{Frontier cost model.}
Since running every candidate would cost more than the query itself, the cost model predicts the runtime of each candidate frontier and picks the candidate with the lowest prediction.
The candidates are no frontier and $H = 512, 1024, 2048, \dots$, which double up to 65,536 values, or up to all values if a relation has fewer.
A smaller frontier moves too few products to the accelerator to pay for its setup, and at 65,536 values the grid's buffers for tile results already take 2\,GB.
For each candidate, the model counts four kinds of work and multiplies each count by a measured time per unit.
The \emph{sparse term} counts the products that involve a value beyond $H$ and, for counts, the work of the heavy-light path.
The \emph{tile term} counts the accelerator's tile products that hold a useful product.
The \emph{setup term} counts the rows and nonzeros that the frontier covers, which \sys must order and pack into tiles.
The \emph{block term} counts the pairs of value blocks, each with its own bookkeeping and results.
One pass over the rows gives the sparse and setup counts of all candidates exactly, because each row's values are in rank order and each candidate is one position in the row.
The tile count is harder, since counting tiles exactly would require building them, which is the very setup that the model must avoid.
Hence, \sys counts the tiles on a sample of a few thousand groups of consecutive rows and scales the count up.
The times per unit are seven constants, which we measure once per machine from a few runs with forced frontiers.

\noindent\textbf{An example of the frontier choice.}
Figure~\ref{fig:frontier} shows the estimate for a text collection.
As $H$ grows, the frequent words move to the grid, and the sparse term falls faster than the tile term grows, since their tiles are full.
Beyond the best frontier, the new values are rare, and their tiles add tile products and setup but remove few sparse products.
\sys picks the minimum, and for a pruned query it recomputes the curve on the remaining values.

\subsection{Feeding the Matrix Accelerator \label{sec:design-grid}}

The matrix accelerator multiplies dense tiles, but even in the frontier, a row holds few of the $H$ values, and most entries of these rows are zero.
A strawman approach would copy the frontier into a dense matrix with one row per key and call a vendor library.
However, the copy would grow with the input rather than with the result, and the library would also multiply the tiles that hold only zeros.
Compressing away the empty columns of each pair of blocks would avoid these tiles, but building and mapping back the compressed operands of every pair would cost as much as it saves.
Our approach is to build each tile directly from the rows that touch it and to skip every tile in which no row holds a value of both blocks.

\begin{figure}[t]
\centering
\input{figures/tikz/fig_grid.tex}
\caption{Example of the grid for one pair of 32-value blocks $b_1$ and $b_2$ over four chunks of 64 rows. \textit{\tsn{1} A bitmap per block marks the chunks that hold its values, and only the chunks common to both are loaded. \tsn{2} In each common chunk, only the halves of the two panels that hold values are multiplied.}\label{fig:grid}}
\vspace{-3mm}
\end{figure}
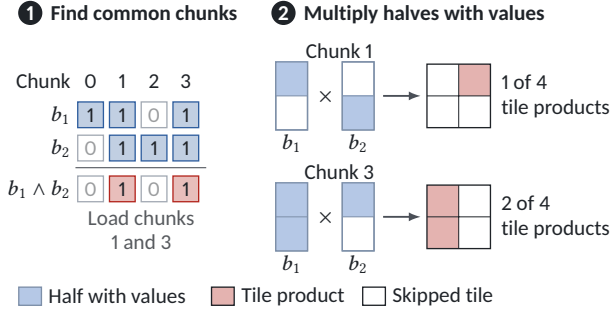

\noindent\textbf{Panels built from rows.}
The accelerator holds only a few tiles at a time, and it multiplies two tiles that it loads from memory into a third that it keeps.
Hence, \sys cuts the dense core into pieces of tile size.
It splits the frontier into \emph{blocks} of values, as wide as the results of a pair of blocks can be while they stay inside the accelerator with the tiles being multiplied, and the rows into \emph{chunks} of consecutive rows.
For each block and chunk that share a nonzero, \sys writes a \emph{panel}, the chunk's entries for the block at one byte each, and marks which halves of the block hold values.
A bitmap per block marks the chunks that have a panel.
Since a panel holds one byte per value, the grid runs counts and weights from 1 to 255.
Larger integer weights run on the sparse path with exact integer sums, and real weights on one with floating-point sums and without pruning.

\noindent\textbf{Row order.}
Rows that hold the same values should share a chunk, since their values then fill a few panels instead of many.
Finding the best order is a hard clustering problem.
Instead, \sys sorts the rows by two MinHash signatures of their frontier values~\cite{broder}, which two rows share with a probability equal to the share of values they have in common.

\noindent\textbf{The pair loop.}
For a pair of blocks $(b_1, b_2)$ with $b_1 \le b_2$, the counts stay inside the accelerator as four result tiles, one for each pair of block halves, and Figure~\ref{fig:grid} shows the two steps that fill them.
First, \sys intersects the chunk bitmaps of the two blocks and reads only the chunks that hold values of both, chunks 1 and 3 in the example (\tsn{1}).
Second, in each common chunk, it multiplies only the panel halves that hold values (\tsn{2}).
As a result, chunk 1 needs one tile product instead of four, and chunk 3 needs two.
Since the result tiles hold 32-bit sums, \sys checks from the column totals of the frontier that no cell can reach $2^{31}$ and otherwise runs the query without a frontier.

\noindent\textbf{Grouped pair loop.}
Each block takes part in many pairs, and the order of the pairs decides how often its panels come from memory.
\sys therefore orders the pairs to reuse panels while they are still in the caches, as a blocked matrix multiplication does.
It takes a group of second blocks $b_2$ and a range of chunks whose panels fit in the CPU's last-level cache.
Within a range, each thread multiplies a few first blocks $b_1$ at a time with every block of the group, which keeps their panels in the core's second-level cache.
Between ranges, the result tiles of a pair wait in memory, and \sys merges consecutive sparse ranges to reduce these reloads.

\noindent\textbf{Heavy-light path.}
The pairs of a frontier value and a rare value are numerous, but each row that holds a rare value also holds many frontier values.
Hence, for counts, the heavy-light path gathers the rows of a batch of rare values and counts their pairs with the frontier values on the matrix accelerator, building the tiles from a bitmap of frontier values per row.

\subsection{Running Inside Database Engines \label{sec:design-engines}}

\sys should run inside existing database engines and follow each engine's own rules for running queries and managing resources, or the engine cannot use it for real queries.
What makes this challenging is that engines differ widely in both respects.
Exporting the input to an external program would sidestep these differences, but it would copy the data out and the result back, and the engine could no longer limit or cancel the work.
Writing the operator inside one engine would avoid the copies, but it would tie the operator to that engine's plan structures, and every further engine would need the whole design again.
Our design keeps the operator independent of any engine and reaches each engine through the interfaces that engines already offer to extensions.
Therefore, the core of \sys is a library with a small C interface, which takes the input rows and returns the result cells, and each engine calls it through a thin adapter.

\noindent\textbf{The operator interface.}
A narrow interface separates the engine and \sys.
It takes the input rows in batches, the query with its options, and a cancellation callback, and it returns result rows that the adapter converts to the engine's types.
Hence, the engine never sees how \sys computes the result, and \sys never sees the engine's plan structures.

\noindent\textbf{What an engine must provide.}
Any engine that offers three common extension points can host \sys.
First, a hook puts \sys in place of the join and the aggregation.
With an extensible optimizer or planner, the adapter replaces the subplan, and users keep writing the original SQL.
Without one, \sys is a table function that the query names.
Second, the adapter reads the input rows with the engine's own scan, and third, it returns the cells as rows in the engine's types.
Since relational, embedded, and dataflow engines all offer these interfaces to extensions, \sys runs in them without changes to their code.

\begin{figure}[t]
\centering
\subfloat[Two-relation product]{\input{figures/tikz/fig_ext_cross.tex}}\hfill
\subfloat[Similarity]{\input{figures/tikz/fig_ext_sim.tex}}
\caption{The same split of the result between the grid and the sparse path serves other queries. \textit{Each square is the result with the values in rank order. (a) A product of two relations has one frontier per input. (b) Item similarity also skips the pairs whose sizes differ too much to reach the threshold.}\label{fig:ext}}
\vspace{-3mm}
\end{figure}
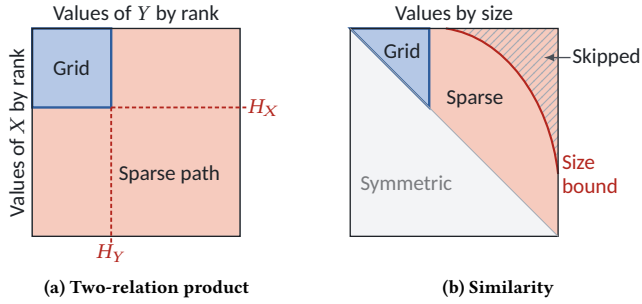

\noindent\textbf{Leaving a query to the engine.}
\sys does not speed up every join, and the adapter leaves the joins that it cannot speed up to the engine.
\sys pays for every input row, because it builds the rows of $X$ from them, while the engine's hash join and aggregation pay for every joined row.
Therefore, a join that adds few rows to its input runs faster in the engine.
The adapter estimates the joined rows per input row from the row and distinct counts that the engine keeps for each table.
Since this estimate assumes uniform keys, it misses the skew that makes a join large and gives the matrix repeated pairs to aggregate.
The adapter therefore leaves a join to the engine only below a limit at which \sys is clearly slower even on uniform keys.

\noindent\textbf{Resources of the engine.}
\sys uses the engine's resources as one of the engine's own operators would.
Its worker threads run on the CPUs of the calling thread and leave signals, such as a statement timeout, to the engine's thread.
The output, which can grow with the square of the number of values, spills to the engine's temporary directory under its memory limit, while the input rows, $X$, and each thread's counters grow linearly and stay in memory.
Since engines cancel cooperatively, \sys polls a cancellation callback between phases and once per work item of the products, and the adapter then raises the engine's own cancellation error.

\subsection{Two-Relation Products and Item Similarity \label{sec:design-ext}}

\sys also supports products of two relations and item similarity.
Both reuse the grid and the sparse path and change only how the ranked result is split between them.

\noindent\textbf{Two-relation products.}
For $X^\top Y$, the values of both inputs share one rank space with four ranges, the frontier of $X$, the frontier of $Y$, and the remaining values of each input.
The grid multiplies only blocks of $X$'s frontier with blocks of $Y$'s frontier, and the sparse path scans only the other input's ranges of each row.
A variant of the cost model chooses the two frontiers together, with the tile term counting tile pairs across the inputs.
Figure~\ref{fig:ext}~(a) shows this split.

\noindent\textbf{Similarity.}
For item similarity, \sys computes each pair's similarity when it emits the cell, from the exact count $C_{i,j}$ and the sizes $C_{i,i}$ and $C_{j,j}$, and keeps the pairs above the threshold $t$.
For sets, a pair can reach $t$ only if the smaller size is at least a fraction of the larger, $t^2$ for cosine and $t$ for Jaccard.
Since values are ranked by size, this becomes a smallest rank for the partners of each value $j$.
The grid skips the blocks below this rank, and the sparse path starts each row's scan at it, as Figure~\ref{fig:ext}~(b) shows.

\subsection{Other Matrix Accelerators \label{sec:design-other}}

Our prototype targets Intel AMX~\cite{intelamx}, and porting it to other CPU matrix accelerators is future work.
Such accelerators also accumulate products of 8-bit values into 32-bit sums, such as the outer-product instructions of Arm SME~\cite{armsme} and the matrix facility of IBM Power10~\cite{power10mma}.
Since the grid and the cost model rely only on such products, we expect them to carry over, with three differences.
First, since the 8-bit products of Power10 take one signed operand, weights above 127 need a second product.
Second, SME tiles follow the vector length of the processor, and the panel width must follow it too.
Third, both accumulate outer products of vectors instead of multiplying two tiles loaded from memory, which changes how panels feed the products and what the cost model charges per tile.

%% file: figures/tikz/fig_overview.tex
\begin{tikzpicture}[x=1cm, y=1cm, font=\figfontS]
  \node[tsregion=tsGray, minimum width=1.35cm, minimum height=4.35cm, anchor=south west] at (0,-0.2) {};
  \node[tstitle, align=left] at (0.07,4.08) {Database\\engine};
  \node[tspill=tsGray, minimum width=0.95cm, align=center] (plan) at (0.5,2.95) {Query\\plan};
  \node[tspill=tsGray, minimum width=0.95cm, align=center] (res) at (0.5,0.75) {Result\\rows};
  \node[draw=tsGray, fill=tsGrayM, line width=0.4pt, minimum width=0.3cm, minimum height=3.35cm, inner sep=0pt] at (1.18,1.6) {};
  \node[tslab, rotate=90] at (1.18,1.6) {Adapter};
  \node[tsregion=tsSlate, minimum width=3.1cm, minimum height=2.0cm, anchor=south west] at (1.65,2.15) {};
  \node[tstitle] at (1.71,4.08) {\tsn{1}\,\tsn{2}\, Rows to ranked $X$};
  \begin{scope}[shift={(1.85,2.74)}, x=0.14cm, y=0.12cm]
    \fill[tsSlateM] (0,2) rectangle (3,4);
    \foreach \r in {0,...,5} \draw[tsSlate, line width=0.35pt] (0,\r) rectangle (3,\r+1);
    \draw[tsSlate, line width=0.35pt] (1.5,0) -- (1.5,6);
    \node[tslab] at (0.75,6.75) {$k$}; \node[tslab] at (2.25,6.75) {$x$};
  \end{scope}
  \node[tslab, anchor=north] at (2.06,2.7) {Rows};
  \draw[tsthin] (2.38,3.1) -- (3.12,3.1);
  \node[tslab, anchor=south] at (2.75,3.11) {Merge};
  \node[tslab, anchor=north] at (2.75,3.09) {rank};
  \begin{scope}[shift={(3.2,2.74)}, x=0.13cm, y=0.12cm]
    \foreach \c/\d in {0/90,1/78,2/66,3/44,4/30,5/20,6/12,7/8} \fill[tsBlue!\d!tsSlateL] (\c,0) rectangle (\c+1,6);
    \draw[tsgrid] (0,0) grid[step=1] (8,6);
    \draw[tsSlate, line width=0.45pt] (0,0) rectangle (8,6);
  \end{scope}
  \node[tslab, anchor=north] at (3.72,2.7) {Frequent first};
  \node[tsregion=tsBlue, minimum width=3.43cm, minimum height=2.0cm, anchor=south west] at (4.95,2.15) {};
  \node[tstitle] at (5.01,4.08) {\tsn{3}\,\tsn{4}\, Prune, choose $H$};
  \begin{scope}[shift={(5.2,2.55)}, x=0.13cm, y=0.13cm]
    \fill[tsCoralM] (0,2) rectangle (6,8);
    \fill[tsBlueM] (0,5) rectangle (3,8);
    \fill[pattern=north east lines, pattern color=tsGray] (6,0) rectangle (8,8);
    \fill[pattern=north east lines, pattern color=tsGray] (0,0) rectangle (6,2);
    \draw[tsgrid] (0,0) grid[step=1] (8,8);
    \draw[tsInk, line width=0.45pt] (0,0) rectangle (8,8);
    \draw[tsBlue, line width=0.9pt] (0,5) rectangle (3,8);
    \draw[tsRed, line width=0.7pt, dash pattern=on 1.6pt off 1.0pt] (3,-0.25) -- (3,8);
    \node[tslab, text=tsRed, anchor=north, inner sep=0.3pt] at (3,-0.2) {$H$};
    \draw[tsInk, line width=0.5pt] (6,-0.25) -- (6,0);
    \node[tslab, anchor=north, inner sep=0.3pt] at (6,-0.2) {$P$};
  \end{scope}
  \node[tslab, anchor=west] (lf) at (6.55,3.45) {Dense core};
  \draw[tsthin] (lf.west) -- (5.45,3.45);
  \node[tslab, anchor=west] (ls) at (6.55,3.03) {Sparse};
  \draw[tsthin] (ls.west) -- (5.9,3.03);
  \node[tslab, anchor=west] (lp) at (6.55,2.61) {Pruned};
  \draw[tsthin] (lp.west) -- (6.12,2.61);
  \node[tsregion=tsBlue, minimum width=4.73cm, minimum height=2.1cm, anchor=south west] at (3.65,-0.2) {};
  \node[tstitle] at (3.71,1.83) {\tsn{5}\, Products, each pair once};
  \node[tsbox=tsCoral, minimum width=2.25cm, minimum height=1.55cm, fill=tsCoralL, anchor=south west] at (3.75,-0.1) {};
  \node[tslab, anchor=north] at (4.875,1.42) {Sparse path, $\ge H$};
  \foreach \y/\a/\b in {1.05/0.25/0.6, 0.85/0.6/0.85} {
    \draw[tsGray, line width=0.35pt] (3.95,\y) -- (5.8,\y);
    \draw[tsRed, line width=0.6pt, -{Latex[length=1.0mm,width=0.8mm]}] (3.95,\y+0.06) -- (5.0,\y+0.06);
    \fill[tsInk] (3.95+\a,\y) circle (1.0pt); \fill[tsInk] (3.95+\b,\y) circle (1.0pt);
    \fill[tsRed] (5.05,\y) circle (1.3pt);
  }
  \foreach \x in {0,...,3} \draw[tsLav, line width=0.35pt] (3.95+0.16*\x,0.42) rectangle ++(0.16,0.2);
  \foreach \x/\n in {0/1,2/2,3/1} {
    \fill[tsLavM] (3.95+0.16*\x,0.42) rectangle ++(0.16,0.2);
    \draw[tsLav, line width=0.45pt] (3.95+0.16*\x,0.42) rectangle ++(0.16,0.2);
  }
  \node[tslab, anchor=west] at (4.64,0.52) {Counters};
  \node[tslab, anchor=north] at (4.875,0.36) {CPU cores};
  \node[tsbox=tsBlue, minimum width=2.28cm, minimum height=1.55cm, fill=tsBlueL, anchor=south west] at (6.05,-0.1) {};
  \node[tslab, anchor=north] at (7.19,1.42) {Grid, $< H$};
  \foreach \x/\up/\dn in {6.53/1/0, 7.03/1/1} {
    \draw[tsBlue, line width=0.4pt, fill=white] (\x,0.45) rectangle ++(0.24,0.6);
    \ifnum\up=1 \fill[tsBlueM] (\x,0.75) rectangle ++(0.24,0.3); \fi
    \ifnum\dn=1 \fill[tsBlueM] (\x,0.45) rectangle ++(0.24,0.3); \fi
    \draw[tsBlue, line width=0.4pt] (\x,0.45) rectangle ++(0.24,0.6); \draw[tsBlue, line width=0.3pt] (\x,0.75) -- ++(0.24,0);
  }
  \node[tslab] at (6.90,0.75) {$\times$};
  \draw[tsthin] (7.35,0.75) -- (7.55,0.75);
  \draw[tsInk, line width=0.4pt] (7.60,0.5) rectangle ++(0.5,0.5);
  \draw[tsInk, line width=0.3pt] (7.85,0.5) -- ++(0,0.5); \draw[tsInk, line width=0.3pt] (7.60,0.75) -- ++(0.5,0);
  \fill[tsRedL] (7.61,0.76) rectangle (7.84,0.99); \fill[tsRedL] (7.86,0.76) rectangle (8.09,0.99);
  \node[tslab, anchor=north] at (7.19,0.36) {Matrix accelerator};
  \node[tsregion=tsLav, minimum width=1.75cm, minimum height=2.1cm, anchor=south west] at (1.65,-0.2) {};
  \node[tstitle] at (1.71,1.83) {\tsn{6}\, Output};
  \foreach \i in {0,1,2} \draw[tsLav, fill=tsLavM, line width=0.45pt] (1.85+\i*0.07,0.95+\i*0.09) rectangle ++(0.62,0.3);
  \draw[tsthin] (2.68,1.15) -- (2.92,1.15);
  \draw[tsLav, fill=white, line width=0.45pt] (2.97,0.98) -- ++(0.28,0) -- ++(0,0.34) -- ++(-0.08,0.08) -- ++(-0.2,0) -- cycle;
  \node[tslab, anchor=north, align=center] at (2.52,0.88) {Per-thread\\chunks, spilled\\to files};
  \draw[tsgo] (plan.east) -- (1.65,2.95);
  \draw[tsgo] (4.75,3.2) -- (4.95,3.2);
  \draw[tsgo] (8.1,2.15) -- (8.1,1.9);
  \draw[tsgo] (3.65,0.68) -- (3.4,0.68);
  \draw[tsgo] (1.65,0.75) -- (res.east);
\end{tikzpicture}

%% file: figures/tikz/fig_sparse.tex
\begin{tikzpicture}[x=0.78cm, y=1cm, font=\figfontS]
  \fill[tsBlueM] (0,2.25) rectangle (1.6,2.55);
  \fill[tsCoralM] (1.6,2.25) rectangle (5.2,2.55);
  \fill[pattern=north east lines, pattern color=tsGray] (5.2,2.25) rectangle (7.4,2.55);
  \draw[tsInk, line width=0.45pt] (0,2.25) rectangle (7.4,2.55);
  \node[tslab] at (0.8,2.4) {Frontier};
  \node[tslab] at (3.4,2.4) {Sparse path};
  \node[tslab, fill=white, inner sep=0.5pt] at (6.3,2.4) {Pruned};
  \node[tslab, anchor=south] at (1.6,2.55) {$H$};
  \node[tslab, anchor=south] at (5.2,2.55) {$P$};
  \node[tstitle, anchor=south west] at (-0.4,2.86) {Values ordered by rank};
  \foreach \k/\y/\pa/\pb/\pc in {1/1.75/2.0/2.8/6.0, 2/1.2/2.8/3.6/6.6} {
    \node[tslab, anchor=east] at (-0.05,\y) {$k_{\k}$};
    \draw[tsGray, line width=0.4pt] (0,\y) -- (7.4,\y);
    \draw[tsRed, line width=0.9pt, -{Latex[length=1.5mm,width=1.1mm]}] (0,\y+0.12) -- (4.15,\y+0.12);
    \foreach \x in {\pa,\pb} \fill[tsInk] (\x,\y) circle (1.6pt);
    \fill[tsRed] (4.2,\y) circle (2.0pt);
    \draw[tsGray, line width=0.6pt, fill=white] (\pc,\y) circle (1.6pt);
  }
  \node[tslab, anchor=south] at (2.0,1.85) {$i_1$};
  \node[tslab, anchor=south] at (2.8,1.85) {$i_2$};
  \node[tslab, anchor=south] at (2.8,1.3) {$i_2$};
  \node[tslab, anchor=south] at (3.6,1.3) {$i_3$};
  \node[tslab, anchor=south, text=tsRed] at (4.2,1.85) {$j$};
  \node[tslab, anchor=south, text=tsRed] at (4.2,1.3) {$j$};
  \node[tslab, anchor=west] at (4.5,1.47) {Scan stops at $j$};
  \foreach \x in {2.0,2.8,3.6} \draw[tsLav, line width=0.5pt, dash pattern=on 1.2pt off 1.0pt] (\x,1.1) -- (\x,0.66);
  \foreach \x in {0.4,0.8,...,4.0} \draw[tsLav, line width=0.4pt] (\x-0.2,0.3) rectangle ++(0.4,0.32);
  \foreach \x/\n in {2.0/1,2.8/2,3.6/1} {
    \fill[tsLavM] (\x-0.2,0.3) rectangle ++(0.4,0.32);
    \draw[tsLav, line width=0.6pt] (\x-0.2,0.3) rectangle ++(0.4,0.32);
    \node[tslab] at (\x,0.46) {\n};
  }
  \node[tslab, anchor=west, align=left] at (4.3,0.46) {Accumulator of column $j$,\\indexed by $i$. Its touched\\entries are the result cells.};
\end{tikzpicture}

%% file: figures/tikz/fig_frontier.tex
\begin{tikzpicture}[x=1cm, y=1.25cm, font=\figfontS]
  \def\ymax{1.8}
  \def\bw{0.42}
  \foreach \t in {0.5,1,1.5} \draw[tsGrayM, line width=0.3pt] (0,\t) -- (6.95,\t);
  \draw[tsArrow, line width=0.5pt] (0,0) -- (6.95,0);
  \draw[tsArrow, line width=0.5pt] (0,0) -- (0,\ymax);
  \node[tslab, rotate=90, anchor=south] at (-0.08,0.9) {Estimated time};
  \node[tslab, anchor=north] at (3.47,-0.25) {Candidate frontier $H$ (values in frequency order)};
  \foreach \k/\l/\sp/\su/\ti/\tot in {0/0/1.000/0.000/0.000/1.00, 1/512/1.174/0.021/0.001/1.20, 2/1K/0.916/0.030/0.005/0.95,
      3/2K/0.704/0.039/0.020/0.76, 4/4K/0.582/0.049/0.078/0.71, 5/8K/0.542/0.056/0.310/0.91, 6/16K/0.515/0.062/1.059/1.64,
      7/32K/0.422/0.066/2.485/2.97, 8/64K/0.195/0.068/4.174/4.44} {
    \pgfmathsetmacro{\x}{0.45+\k*0.76}
    \pgfmathsetmacro{\a}{\sp}
    \pgfmathsetmacro{\b}{\sp+\su}
    \pgfmathsetmacro{\c}{min(\sp+\su+\ti,\ymax)}
    \filldraw[fill=tsCoralM, draw=tsCoral, line width=0.4pt] (\x-\bw/2,0) rectangle (\x+\bw/2,\a);
    \ifdim\su pt>0pt
      \filldraw[fill=tsGrayM, draw=tsArrow, line width=0.4pt] (\x-\bw/2,\a) rectangle (\x+\bw/2,\b);
    \fi
    \ifdim\ti pt>0pt
      \filldraw[fill=tsBlueM, draw=tsBlue, line width=0.4pt] (\x-\bw/2,\b) rectangle (\x+\bw/2,\c);
    \fi
    \ifdim\tot pt>\ymax pt
      \draw[white, line width=1.1pt] (\x-\bw/2-0.03,\ymax-0.17) -- (\x+\bw/2+0.03,\ymax-0.11);
    \fi
    \node[tslab, anchor=north] at (\x,-0.02) {\l};
  }
  \draw[tsRed, line width=0.7pt, dash pattern=on 1.6pt off 1.0pt] (3.49-0.27,0) rectangle (3.49+0.27,0.75);
  \node[tslab, text=tsRed, anchor=south, align=center, inner sep=1pt] at (3.49,0.77) {Chosen\\[-1pt]$H$};
  \begin{scope}[shift={(1.85,1.73)}]
    \foreach \n/\f/\d/\t in {0/tsCoralM/tsCoral/Sparse term, 1/tsGrayM/tsArrow/Setup term, 2/tsBlueM/tsBlue/Tile and block terms} {
      \filldraw[fill=\f, draw=\d, line width=0.4pt] (0,-\n*0.2-0.05) rectangle (0.2,-\n*0.2+0.05);
      \node[tslab, anchor=west] at (0.24,-\n*0.2) {\t};
    }
  \end{scope}
\end{tikzpicture}

%% file: figures/tikz/fig_grid.tex
\begin{tikzpicture}[x=1cm, y=1cm, font=\figfontS]
  \node[tstitle] at (-0.05,3.35) {\tsn{1}\, Find common chunks};
  \node[tslab, anchor=east] at (0.72,2.25) {Chunk};
  \foreach \c in {0,...,3} \node[tslab] at (0.95+\c*0.42,2.25) {\c};
  \foreach \lab/\y/\bits/\col in {$b_1$/1.8/{1,1,0,1}/tsBlue, $b_2$/1.36/{0,1,1,1}/tsBlue, $b_1 \wedge b_2$/0.82/{0,1,0,1}/tsRed} {
    \node[tslab, anchor=east] at (0.72,\y) {\lab};
    \foreach \b [count=\c from 0] in \bits {
      \ifnum\b=1 \draw[\col, fill=\col!30, line width=0.5pt] (0.78+\c*0.42,\y-0.17) rectangle ++(0.34,0.34);
      \node[tslab] at (0.95+\c*0.42,\y) {1};
      \else \draw[tsGray, fill=white, line width=0.5pt] (0.78+\c*0.42,\y-0.17) rectangle ++(0.34,0.34);
      \node[tslab, text=tsInk!45] at (0.95+\c*0.42,\y) {0}; \fi
    }
  }
  \draw[tsArrow, line width=0.4pt] (0.75,1.1) -- (2.48,1.1);
  \node[tslab, align=center, text=tsInk!75] at (1.62,0.28) {Load chunks\\1 and 3};
  \newcommand{\panel}[5]{%
    \draw[tsSlate, line width=0.5pt] (#1,#2) rectangle ++(0.42,0.9);
    \ifnum#4=1 \fill[tsBlueM] (#1,#2+0.45) rectangle ++(0.42,0.45); \fi
    \ifnum#5=1 \fill[tsBlueM] (#1,#2) rectangle ++(0.42,0.45); \fi
    \draw[tsSlate, line width=0.5pt] (#1,#2) rectangle ++(0.42,0.9);
    \draw[tsSlate, line width=0.35pt] (#1,#2+0.45) -- ++(0.42,0);
    \node[tslab, anchor=north] at (#1+0.21,#2-0.02) {#3};}
  \newcommand{\ctile}[6]{%
    \foreach \v/\dx/\dy in {#3/0/0.42,#4/0.42/0.42,#5/0/0,#6/0.42/0} {
      \ifnum\v=1 \fill[tsRed!35] (#1+\dx,#2+\dy) rectangle ++(0.42,0.42); \fi
    }
    \draw[tsInk, line width=0.5pt] (#1,#2) rectangle ++(0.84,0.84);
    \draw[tsInk, line width=0.35pt] (#1+0.42,#2) -- ++(0,0.84) (#1,#2+0.42) -- ++(0.84,0);}
  \foreach \ch/\y/\ua/\la/\ub/\lb/\tl/\tr/\bl/\br/\n in {1/1.6/1/0/0/1/0/1/0/0/1, 3/0.0/1/1/1/0/1/0/1/0/2} {
    \node[tslab, anchor=south] at (4.24,\y+0.92) {Chunk \ch};
    \panel{3.4}{\y}{$b_1$}{\ua}{\la}
    \node[tslab] at (4.06,\y+0.45) {$\times$};
    \panel{4.28}{\y}{$b_2$}{\ub}{\lb}
    \draw[tsgo] (4.82,\y+0.45) -- (5.32,\y+0.45);
    \ctile{5.4}{\y+0.03}{\tl}{\tr}{\bl}{\br}
    \node[tslab, anchor=west, align=left] at (6.35,\y+0.45) {\n{} of 4\\tile products};
  }
  \node[tstitle] at (3.3,3.35) {\tsn{2}\, Multiply halves with values};
  \foreach \x/\fc/\dc/\lab in {0.0/tsBlueM/tsSlate/Half with values, 2.55/tsRed!35/tsInk/Tile product, 4.55/white/tsInk/Skipped tile} {
    \fill[\fc] (\x,-0.72) rectangle ++(0.3,0.24); \draw[\dc, line width=0.5pt] (\x,-0.72) rectangle ++(0.3,0.24);
    \node[tslab, anchor=west] at (\x+0.36,-0.6) {\lab};
  }
\end{tikzpicture}

%% file: figures/tikz/fig_ext_cross.tex
\begin{tikzpicture}[x=2.75cm, y=2.75cm, font=\figfontS]
  \fill[tsCoralM] (0,0) rectangle (1,1);
  \fill[tsBlueM] (0,0.62) rectangle (0.38,1);
  \draw[tsgrid] (0,0) rectangle (1,1);
  \draw[tsBlue, line width=0.9pt] (0,0.62) rectangle (0.38,1);
  \draw[tsInk, line width=0.5pt] (0,0) rectangle (1,1);
  \draw[tsRed, line width=0.7pt, dash pattern=on 1.6pt off 1.0pt] (0.38,-0.03) -- (0.38,0.62);
  \draw[tsRed, line width=0.7pt, dash pattern=on 1.6pt off 1.0pt] (0.38,0.62) -- (1.03,0.62);
  \node[tslab, text=tsRed, anchor=north, inner sep=0.5pt] at (0.38,-0.03) {$H_Y$};
  \node[tslab, text=tsRed, anchor=west, inner sep=0.5pt] at (1.03,0.62) {$H_X$};
  \node[tslab] at (0.19,0.81) {Grid};
  \node[tslab] at (0.66,0.3) {Sparse path};
  \node[tslab, anchor=south] at (0.5,1.01) {Values of $Y$ by rank};
  \node[tslab, anchor=south, rotate=90] at (-0.01,0.5) {Values of $X$ by rank};
  \path (0.5,-0.15);
\end{tikzpicture}

%% file: figures/tikz/fig_ext_sim.tex
\begin{tikzpicture}[x=2.75cm, y=2.75cm, font=\figfontS]
  \fill[tsGrayL] (0,0) -- (0,1) -- (1,0) -- cycle;
  \fill[tsCoralM] (0,1) -- (1,1) -- (1,0) -- cycle;
  \fill[tsBlueM] (0,1) -- (0.38,1) -- (0.38,0.62) -- cycle;
  \fill[pattern=north east lines, pattern color=tsGray] (0.46,1) .. controls (0.74,0.96) and (0.96,0.66) .. (1,0.3)
    -- (1,1) -- cycle;
  \draw[tsBlue, line width=0.9pt] (0,1) -- (0.38,1) -- (0.38,0.62) -- cycle;
  \draw[tsRed, line width=0.8pt] (0.46,1) .. controls (0.74,0.96) and (0.96,0.66) .. (1,0.3);
  \draw[tsInk, line width=0.5pt] (0,0) rectangle (1,1);
  \draw[tsgrid] (0,1) -- (1,0);
  \node[tslab] at (0.25,0.89) {Grid};
  \node[tslab] at (0.6,0.66) {Sparse};
  \node[tslab, anchor=west, inner sep=0.5pt] (sk) at (1.06,0.86) {Skipped};
  \draw[tsthin] (sk.west) -- (0.93,0.86);
  \node[tslab, text=tsInk!60] at (0.27,0.25) {Symmetric};
  \node[tslab, text=tsRed, anchor=west, align=left, inner sep=0.5pt] at (1.01,0.3) {Size\\bound};
  \node[tslab, anchor=south] at (0.5,1.01) {Values by size};
  \path (0.5,-0.15);
\end{tikzpicture}

%% file: sections/implementation.tex
\begin{table}[t]
\centering\small
\caption{How the adapters attach \sys to four engines, and where users install it.\label{tab:engines}}
\setlength{\tabcolsep}{2.5pt}
\begin{tabular}{@{}lllll@{}}
\toprule
Engine & Plan hook & Input rows & Package & Lines \\
\midrule
\duckdb~v2.0.0 & optimizer rule & parallel sink & community & 2,016 \\
\pg~18.6 & planner hook & heap scan & PGXN & 1,052 \\
\spark~4.2.0 & planner strategy & partitions & Maven Central & 1,050 \\
\sqlite~3.54.1 & table function & statement & PyPI & 526 \\
\bottomrule
\end{tabular}
\end{table}
\begin{table}[t]
\centering\small
\caption{Relations $R(k, x)$ in the evaluation.\label{tab:datasets}}
\setlength{\tabcolsep}{4pt}
\begin{tabular}{@{}llrrr@{}}
\toprule
Relation & $k$ / $x$ & Keys & Values & Rows \\
\midrule
NYTimes~\cite{uciBoW} & Article / word & 300\,K & 102\,K & 69.7\,M \\
PubMed~\cite{uciBoW} & Abstract / word & 8.2\,M & 141\,K & 483\,M \\
MovieLens~\cite{movielens} & User / movie & 201\,K & 84\,K & 32.0\,M \\
kosarak~\cite{fimi} & Session / page & 990\,K & 41\,K & 8.0\,M \\
human\_gene1~\cite{suitesparse} & Gene / gene & 22\,K & 22\,K & 24.7\,M \\
LiveJournal~\cite{snap,comlj} & Vertex / vertex & 4.0\,M & 4.0\,M & 69.4\,M \\
Orkut~\cite{snap,comlj} & Vertex / vertex & 3.1\,M & 3.1\,M & 234\,M \\
\bottomrule
\end{tabular}
\end{table}

\begin{figure*}[t]
\centering
\includegraphics[width=0.8\textwidth]{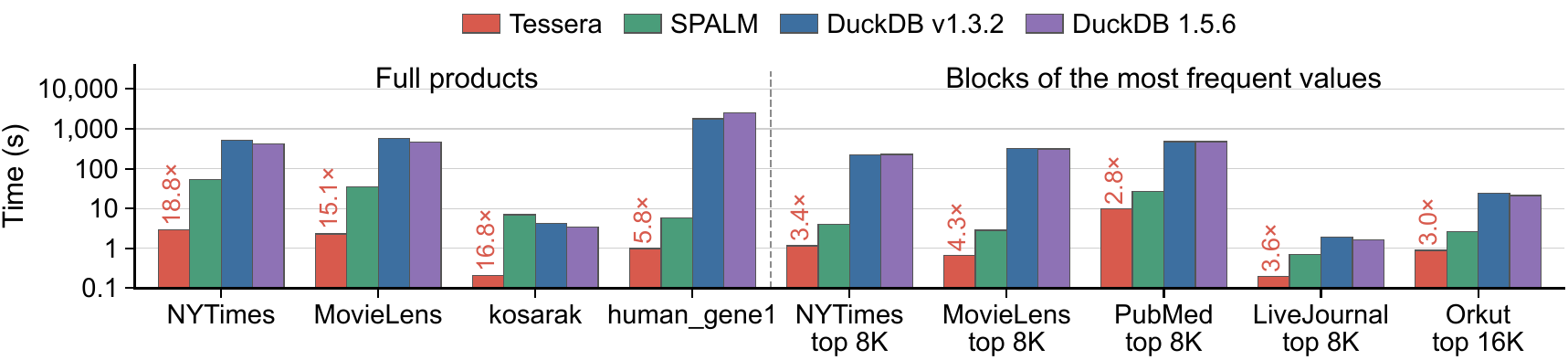}
\caption{Execution time in \duckdb. \sys is faster than every existing system on every relation. \textit{Labels give the speedup over the fastest existing system.}\label{fig:e1}}
\end{figure*}
\begin{figure*}[t]
\centering
\includegraphics[scale=0.5]{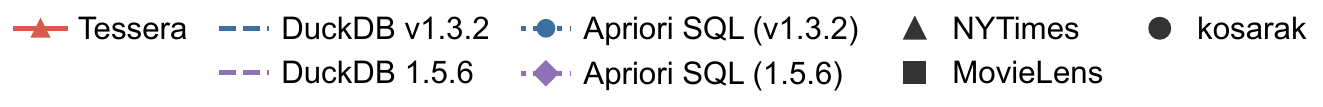}\\
\subfloat[NYTimes, \texttt{HAVING}]{\includegraphics[width=0.24\textwidth]{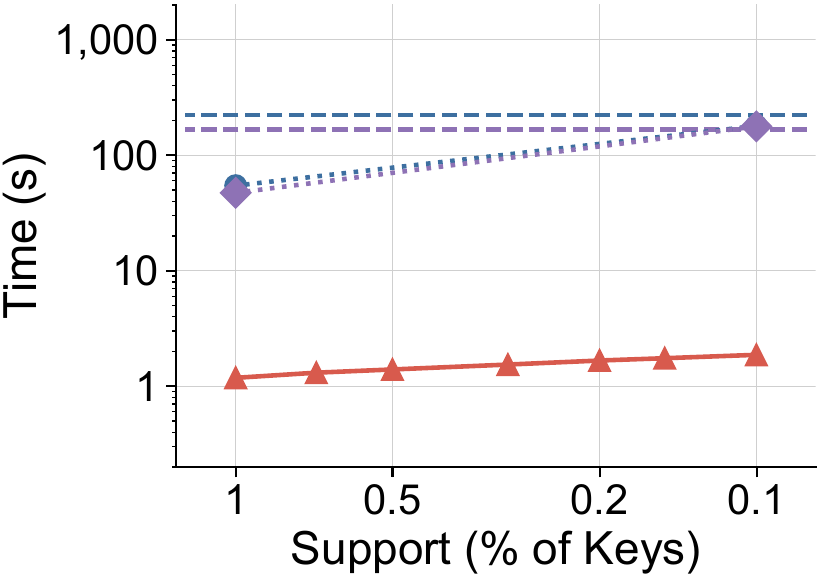}}\hfill
\subfloat[MovieLens, \texttt{HAVING}]{\includegraphics[width=0.24\textwidth]{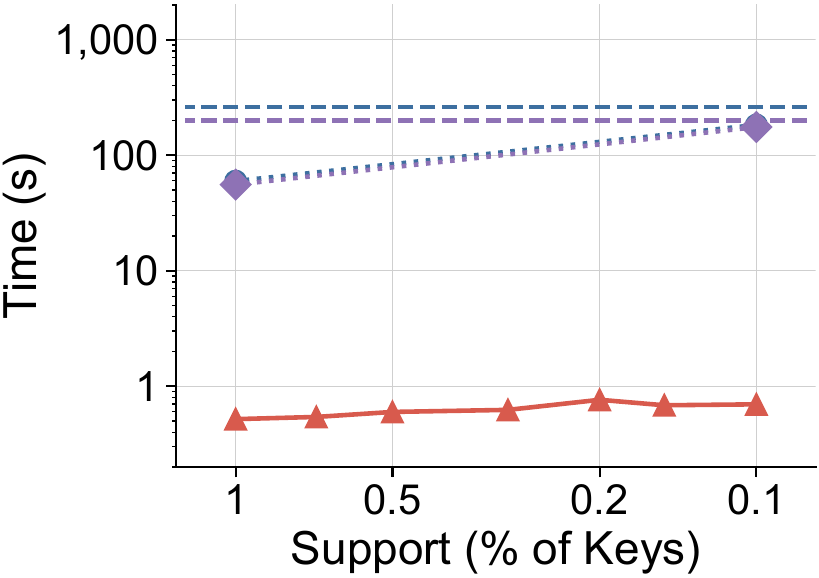}}\hfill
\subfloat[kosarak, \texttt{HAVING}]{\includegraphics[width=0.24\textwidth]{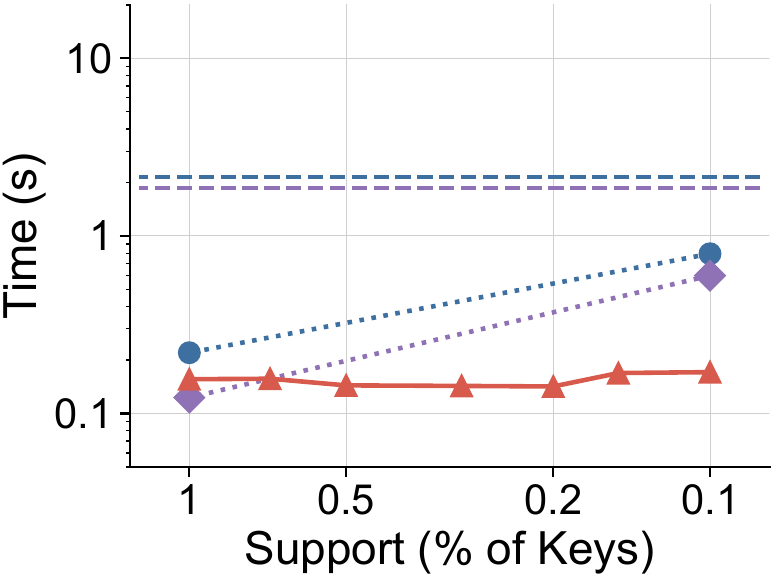}}\hfill
\subfloat[Top-$k$]{\includegraphics[width=0.24\textwidth]{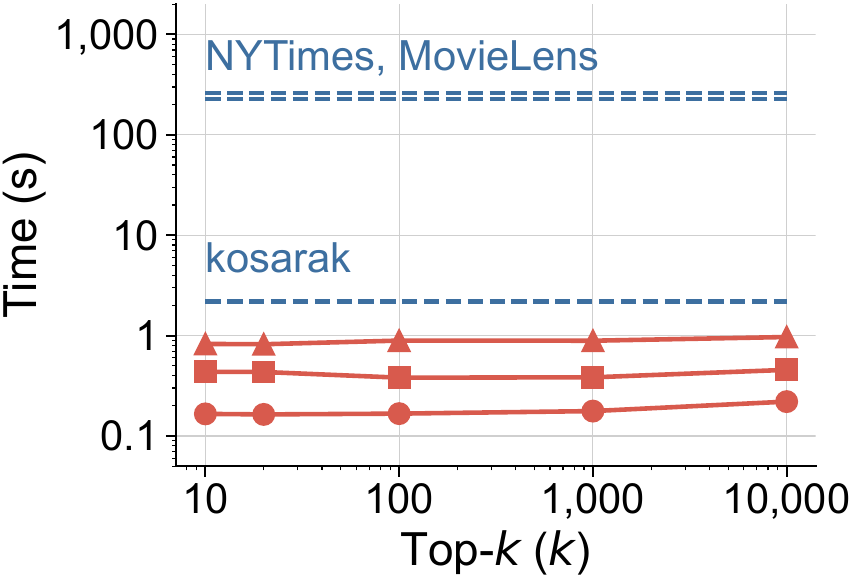}}
\caption{Execution time of pruned queries in \duckdb, which keep pairs above a support (a--c) or the top $k$ (d). \sys is far faster than the engine at every support and $k$, since it drops the values that cannot qualify before counting. \textit{Support is the share of keys in which a pair must appear.}\label{fig:e2}}
\end{figure*}

\section{Implementation \label{sec:impl}}
\sys is released for four widely used database engines, \duckdb, \pg, \spark, and \sqlite, and users install it with one command, as they install any other extension of their engine.
It consists of a core library of about 4,600 lines of C++ and one adapter per engine, built with the core into an extension that changes no engine source (Table~\ref{tab:engines}).
In \duckdb, \pg, and \spark, the adapter rewrites the plan of the original SQL, which needs no change, and in \sqlite, \sys runs as table functions.
On x86-64, the core picks AMX, AVX-512, or AVX2 at runtime, and other CPUs, such as Apple silicon, run a portable build.
Without AMX, every value runs on the sparse path, and on an x86-64 CPU without AVX2, the adapters leave every query to the engine.
The grid uses the AMX instruction \texttt{TDPBUUD} for products of unsigned 8-bit values.

\noindent\textbf{Distribution.}
In \duckdb, \texttt{INSTALL tessera FROM community} fetches the extension, which the community extension repository builds from the public source.
\pg users run \texttt{pgxn install tessera\_pg}, which builds the extension against the local server, and \spark users add the artifact on Maven Central with \texttt{--packages}.
Python programs install the \sqlite extension from PyPI, and other \sqlite users load it from a release archive.
The binary packages target Linux on x86-64 and macOS on Apple silicon.
At each release, a continuous-integration workflow builds and tests the \pg, \spark, and \sqlite packages.

\noindent\textbf{The C interface.}
The \pg, \sqlite, and \spark adapters call the C interface of the core library, and the \duckdb adapter calls the same query code in C++.
An adapter passes the rows of its engine's scan, the memory limit, and a cancellation callback, and reads back the result rows.
Strings are coded in byte order, which keeps the meaning of \texttt{a.x < b.x} under a byte-order collation.

\noindent\textbf{Adapters.}
The adapters of \duckdb, \pg, and \spark leave a join to the engine below a limit on its estimated joined rows per input row.
We set each limit where \sys becomes twice as slow as the engine's plan for a self-join on uniform keys, which gives six in \duckdb~v2.0.0, four in the v1.3.2 build of the evaluation, and no limit in \pg and \spark, whose plans are never twice as fast.
Both builds make the same decisions and take 2.9 seconds on the news articles.
The adapters pass thresholds and the engine's memory limit for the output of \sys, except the \spark adapter, which runs \sys in the driver.
On integer and string keys, weights of one type, pair filters, thresholds, two relations, and similarity, every adapter returns the same rows as the engine's own plan, up to floating-point rounding, and raises the same overflow errors.

%% file: sections/evaluation.tex
\section{Evaluation \label{sec:eval}}

\subsection{Methodology \label{sec:eval-method}}

\noindent\textbf{Testbed.}
The server has two Intel Xeon Silver 4514Y processors with 16 cores, 2\,MB of L2 cache per core, and 30\,MB of L3 cache each, and 256\,GB of DDR5 memory~\cite{intel4514y}, and it runs Ubuntu~24.04 with Linux~6.17.
Every run uses the 16 cores of one socket and the 128\,GB of its NUMA node, and every engine runs with the same 100\,GB memory limit, which keeps it within this node.
Each time is the median of five runs after a warm-up, and runs longer than a minute are measured once.

\noindent\textbf{Compared systems.}
We compare \sys with the engines' own plans, with \spalm, and with programs outside the engines, all from their public code without changes.
The \emph{engines' own plans} build every joined row and then aggregate the joined rows.
In \duckdb, we run the plans of \duckdb~v1.3.2, which also hosts \sys, and of the newer \duckdb~1.5.6.
\spalm~\cite{spalm} picks a sparse or a dense kernel for each block of the product inside \duckdb~v1.3.2 and writes a dense result of all pairs.
It runs with the constants of its own calibration script and per block of result rows where its 32-bit offsets overflow.
Since it does not accept \texttt{a.x < b.x}, it is left out of the pruned queries.
On each \duckdb query, we call the fastest of the two plans and \spalm the \emph{fastest existing system}.
Inside the engines, the times include the engine's scan and the output in the engine's types.

\noindent\textbf{Programs outside the engines.}
The core of \sys runs on rows already in memory against these programs.
The SYRK of oneMKL~\cite{onemkl} and SuiteSparse:\grb~\cite{graphblas,graphblas2} are sparse matrix libraries that compute $X^\top X$, and MMJoin~\cite{deepjoinproject,mmjoin} implements the join-project algorithm.
Eclat~\cite{borgelteclat} and FP-growth~\cite{borgeltfpgrowth} mine frequent itemsets, which we limit to size two.
AllPairs~\cite{allpairs}, PPJoin~\cite{ppjoin}, and GroupJoin~\cite{groupjoin} from the benchmark of Mann et al.~\cite{ssjbench}, the parallel AllPairs of Fier et al.~\cite{fiermulticore}, and L2AP~\cite{l2ap} find similar sets with prefix filters.
Single-thread programs run against one thread of \sys.

\begin{figure}[t]
\centering
\includegraphics[width=\linewidth]{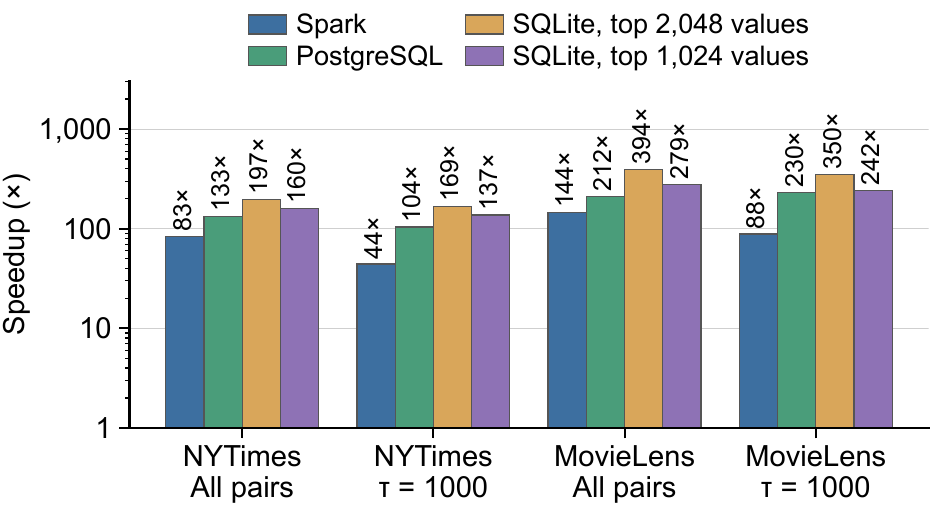}
\caption{\sys outperforms the plans of \spark, \pg, and \sqlite on every query. \textit{All pairs is the full product of the block, and $\tau$ is the \texttt{HAVING} threshold.}\label{fig:g1}}
\end{figure}

\noindent\textbf{Workloads.}
Table~\ref{tab:datasets} lists the relations.
With values numbered by frequency, \texttt{x < 8192} selects the block of the 8,192 most frequent values, the heavy block that motivates in-database linear algebra~\cite{spalm,deepjoinproject}.
For PubMed and the graphs, we run only such blocks, since a dense result of all pairs, which \spalm writes, would need 148\,GB on PubMed and terabytes on the graphs.
We run the full product with \texttt{COUNT} and two \emph{pruned queries}, which keep only strong pairs.
The first is \texttt{a.x < b.x} with a \texttt{HAVING} threshold from 1\% to 0.1\% \emph{support}, the share of keys in which a pair must appear~\cite{fimi}, and the second is top-$k$ with $k$ from 10 to $10^4$.
The deep dive uses the threshold $\tau = 1000$, which is 0.33\% support on the news articles and 0.50\% on the ratings.
Two-relation products join the movie ratings with the 2.0 million tags that users attach to movies, once on the user and once on the movie.
They also join the cast, the keywords, and the companies of IMDb movies on the movie.
Similarity compares the items' sets of keys, and a light consumer computes the count and the sum of the result rows.

\subsection{Main Results \label{sec:eval-main}}

\noindent\textbf{Comparison with existing systems.}
Figure~\ref{fig:e1} shows that \sys outperforms every existing system on every relation.
On the full products, it is faster than the fastest existing system by 5.8$\times$ to 18.8$\times$, and on the heavy blocks, which suit \spalm best, by 2.8$\times$ to 4.3$\times$.
For example, on the news articles, \sys takes 2.9 seconds, \spalm 54.3 seconds, and the engine 8.6 minutes.
On the gene matrix, the engine takes 30 minutes against 1.0 second for \sys, a gain of 1,802$\times$.
This is because the engine processes every joined row and \spalm zeroes and scans a dense result of all pairs, while \sys computes one half of the result and writes only nonzero cells.

\noindent\textbf{Impact of the query's threshold.}
Figure~\ref{fig:e2} shows that the time of \sys grows only slowly as a pruned query keeps more pairs, while the engine takes the same time at every threshold and $k$.
For example, its top-$k$ queries on the news articles take under one second for every $k$, against 226 seconds for the engine.
This is because the engine filters the aggregated groups, while the bound of \sys removes values before the product.
\sys also outperforms the frequent-item filter of Apriori~\cite{apriori,sarawagiarm} written in SQL, which gives the engine part of this bound, by 40$\times$ to 252$\times$ on the news articles and the ratings.
Only on kosarak at 1\% support, where few values reach the threshold, is the rewrite faster, by 1.3$\times$.

\begin{figure}[t]
\centering
\includegraphics[scale=0.5]{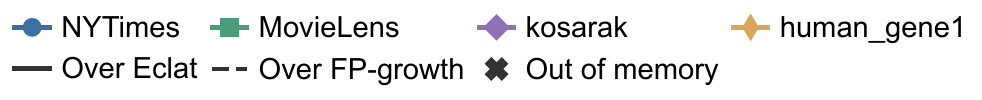}\\
\subfloat[Against sparse libraries]{\includegraphics[width=0.48\linewidth]{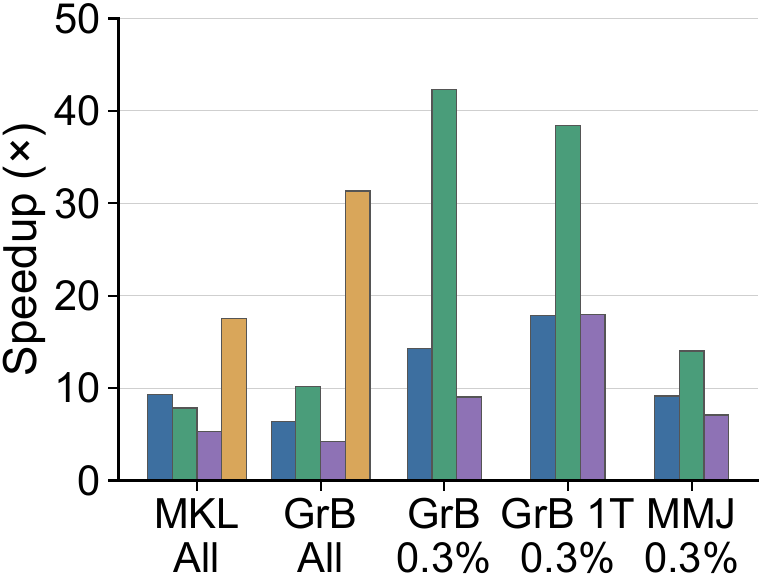}}\hfill
\subfloat[Frequent pairs, one thread]{\includegraphics[width=0.48\linewidth]{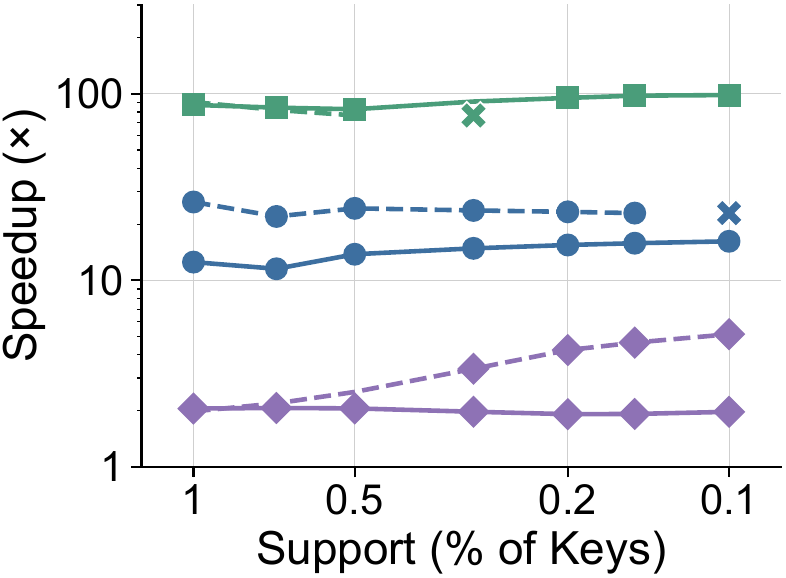}}
\caption{Speedup of the core of \sys outside the engines over sparse libraries, MMJoin, and itemset miners. \textit{MKL is oneMKL SYRK, GrB is \grb, MMJ is MMJoin, All is all pairs, and 1T is one thread.}\label{fig:e3}}
\end{figure}

\begin{figure}[t]
\centering
\includegraphics[scale=0.5]{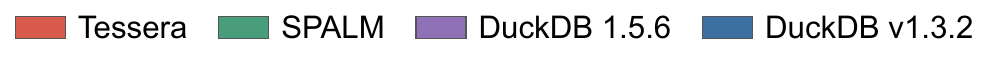}\\
\subfloat[MovieLens ratings $\times$ tags]{\includegraphics[width=0.48\linewidth]{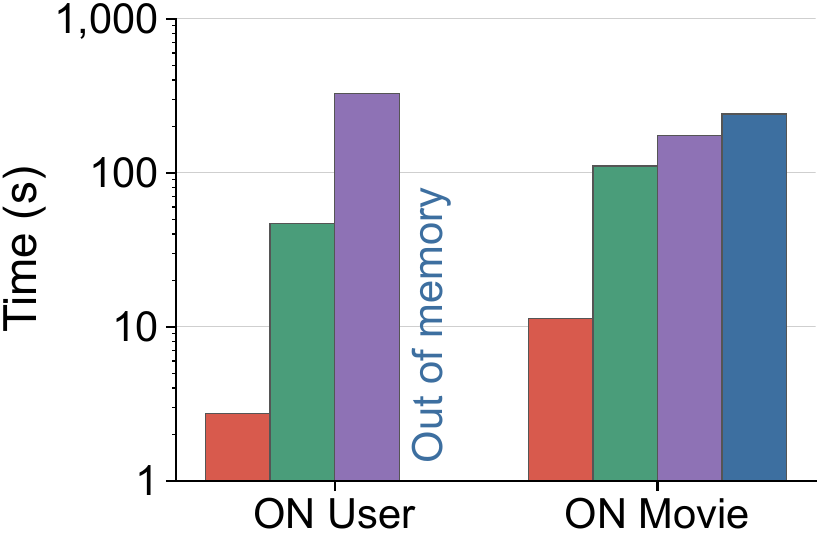}}\hfill
\subfloat[IMDb, joined on the movie]{\includegraphics[width=0.48\linewidth]{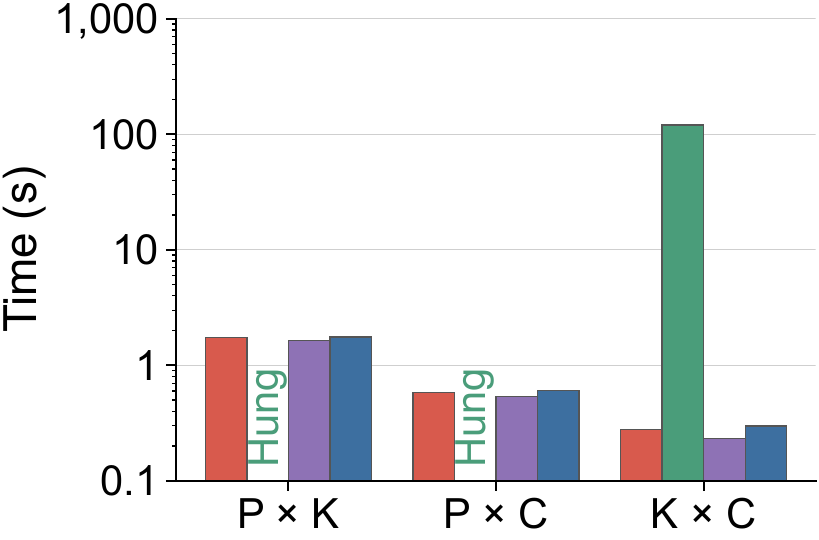}}
\caption{Products of two relations in \duckdb. \sys speeds up joins with many joined rows and leaves the others to the engine. \textit{P, K, and C are the cast, the keywords, and the companies.}\label{fig:x1}}
\end{figure}

\begin{figure*}[t]
\centering
\includegraphics[scale=0.5]{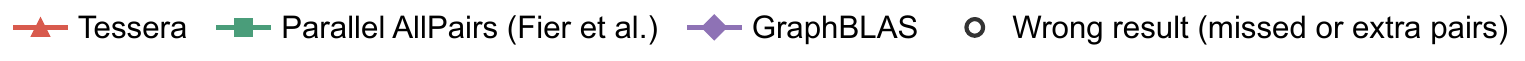}\\
\subfloat[kosarak, Jaccard]{\includegraphics[width=0.24\textwidth]{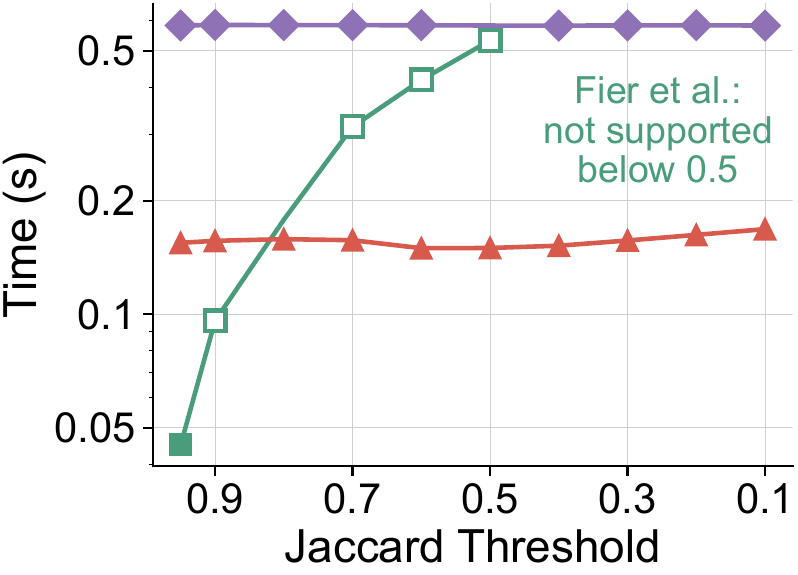}}\hfill
\subfloat[kosarak, cosine]{\includegraphics[width=0.24\textwidth]{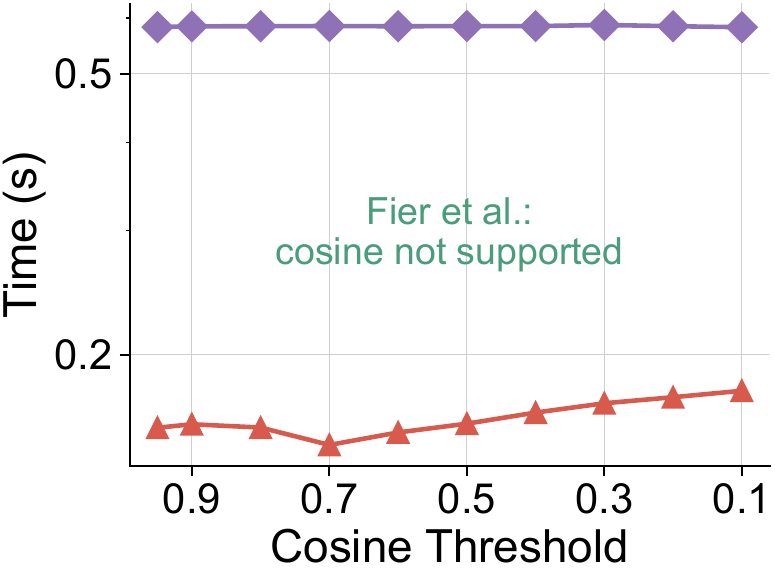}}\hfill
\subfloat[MovieLens, Jaccard]{\includegraphics[width=0.24\textwidth]{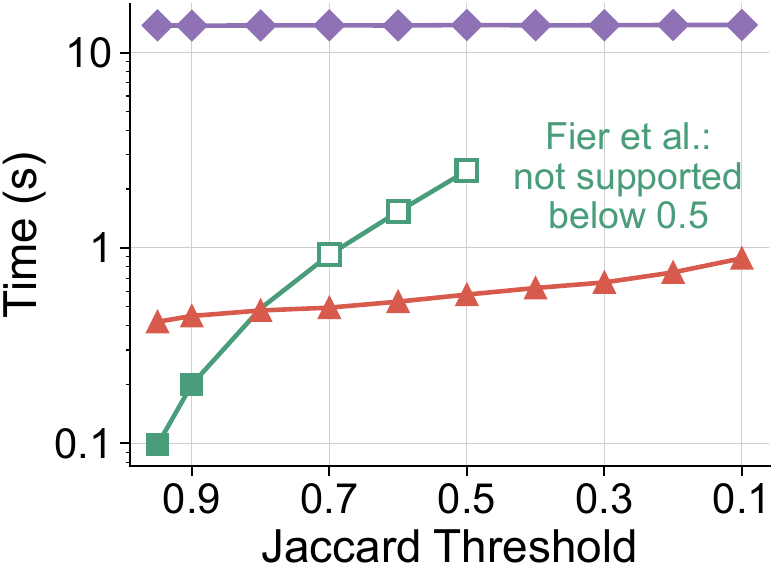}}\hfill
\subfloat[MovieLens, cosine]{\includegraphics[width=0.24\textwidth]{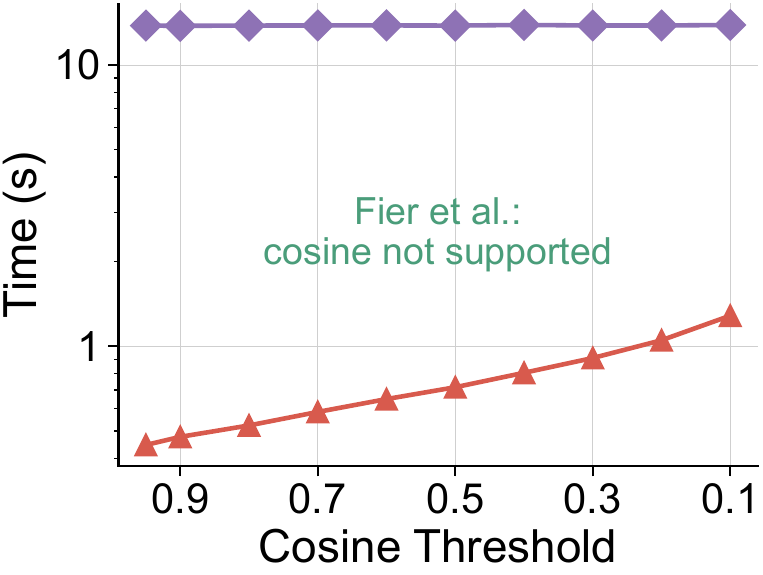}}
\caption{Execution time of item similarity joins outside the engines on 16 threads. \sys is faster than \grb at every threshold and than the parallel AllPairs of Fier et al.\ at 0.7 and below, with exact results. \textit{Hollow markers are runs that miss pairs or return extra ones.}\label{fig:sim16}}
\end{figure*}

\noindent\textbf{Speedup in other engines.}
In \pg and \spark, the queries run on the block of the 8,192 most frequent values.
In \sqlite, whose plan runs on one thread, they run on the blocks of the 2,048 and 1,024 most frequent values.
Figure~\ref{fig:g1} shows that \sys outperforms the plans of \spark, \pg, and \sqlite on every query, by 44$\times$ to 394$\times$.
This is because these plans also build and aggregate every joined row.
Even on one thread, \sys is faster than the plan of \sqlite by 95$\times$ to 253$\times$.

\noindent\textbf{Comparison with sparse libraries and itemset miners.}
Figure~\ref{fig:e3}~(a) shows that the core of \sys outperforms the SYRK of oneMKL and \grb on the full product by 4.2$\times$ to 31.3$\times$.
This is because both assemble a sparse matrix, and \grb also computes both halves and applies the threshold after the product.
The core of \sys, which writes the pairs, is also faster than MMJoin, which only counts them, by 7.1$\times$ to 14.0$\times$ at 0.3\% support.
Figure~\ref{fig:e3}~(b) shows that on one thread, \sys counts the frequent pairs faster than Eclat and FP-growth by 11.5$\times$ to 98.6$\times$ on the news articles and the ratings, where FP-growth does not run out of memory.
On kosarak, whose sessions are short, this gain shrinks to 1.9--5.1$\times$.

\subsection{Two Relations and Item Similarity \label{sec:eval-ext}}

We now show that \sys also supports two other kinds of join-aggregates, products of two relations and item similarity, and that it speeds them up where they produce many joined rows or result pairs.

\noindent\textbf{Two-relation products.}
Figure~\ref{fig:x1}~(a) shows that \sys is faster than \spalm by 9.8$\times$ to 17.1$\times$ and than \duckdb~1.5.6 by 15.3$\times$ to 120.3$\times$ on the two joins, while \duckdb~v1.3.2 runs out of memory on the join on the user.
Since the tags have no dense core, the cost model chooses no frontier, and the gain comes from the sparse path and the output.
Outside the engine, the core of \sys still outperforms \grb by 1.17$\times$ to 1.89$\times$.

\noindent\textbf{Joins that add few rows.}
Figure~\ref{fig:x1}~(b) shows that the adapter rightly leaves joins with few joined rows to the engine.
The IMDb joins give only 2.1--5.3 joined rows per input row, and the estimate from the statistics of \duckdb, 1.1--2.0, lies below the limit of four.
On these joins, \duckdb~1.5.6 outperforms a forced \sys by 1.20$\times$ to 3.40$\times$, since building $X$ from up to 39 million rows takes longer than the engine's whole plan.
With the limit, the three queries run in the engine's own time, while \spalm stops on two of the three.

\begin{figure}[t]
\centering
\includegraphics[scale=0.5]{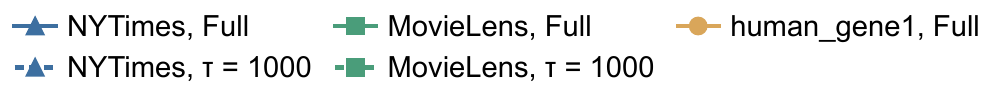}\\
\subfloat[Gain of the matrix accelerator]{\includegraphics[width=0.48\linewidth]{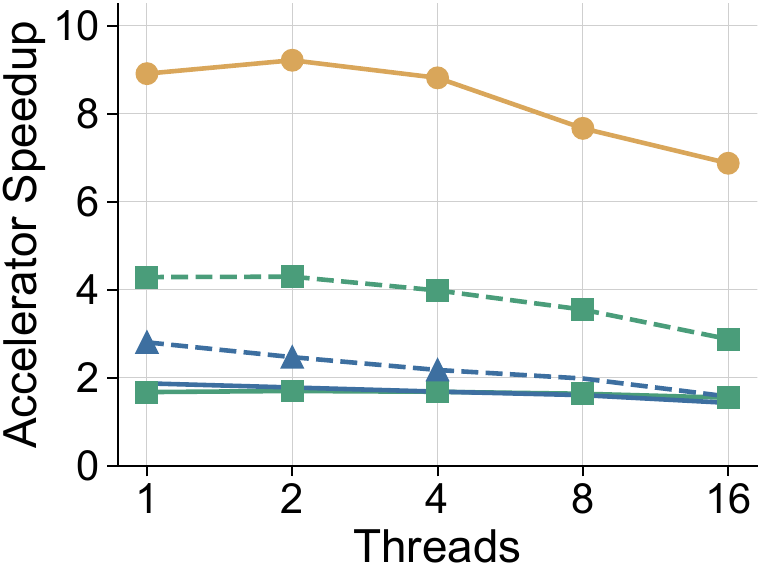}}\hfill
\subfloat[Scaling of \sys]{\includegraphics[width=0.48\linewidth]{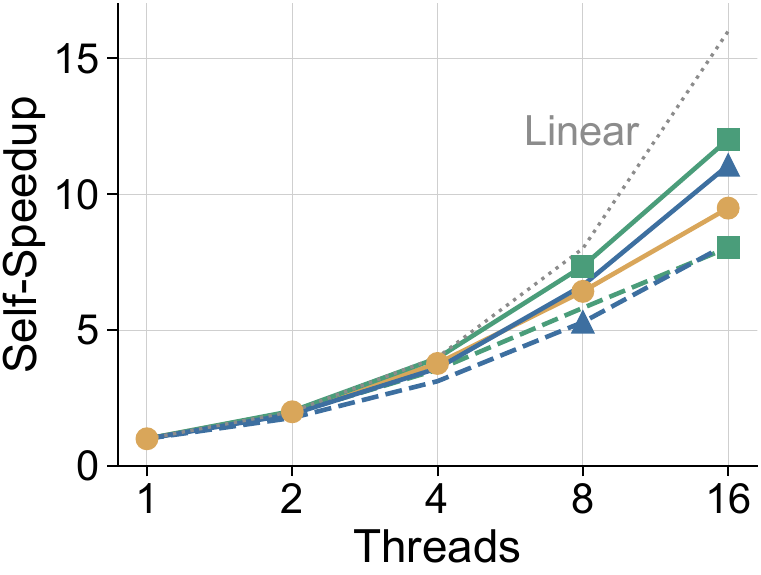}}
\caption{The matrix accelerator speeds up the core most where values form a dense block, and \sys scales with the number of threads.\label{fig:d1}}
\end{figure}

\noindent\textbf{Similarity joins.}
Figure~\ref{fig:sim16} shows that the time of \sys grows by at most 2.9$\times$ as the threshold falls from 0.95 to 0.1.
On the ratings with cosine similarity, the result grows from 1.6 million to 222 million pairs over this range, while the time of \sys grows only from 0.45 to 1.29 seconds.
On 16 threads, \sys outperforms \grb, which filters after the full product, by 3.3$\times$ to 32.9$\times$, and the parallel AllPairs of Fier et al.\ by up to 4.3$\times$ at 0.7 and below.
On one thread, the joins of Mann et al.\ and L2AP lead by up to 75$\times$ at 0.95, while \sys outperforms them by 2.3$\times$ to 3.2$\times$ at 0.1 on the news articles and the ratings.
This is because a prefix filter prunes fewer candidates as the threshold falls, while \sys counts each pair that shares a key once at any threshold.
\sys also returns the exact pairs, while the parallel join compares floating-point bounds with the threshold and misses up to 4.9\% of the result.

\subsection{Deep Dive \label{sec:eval-deep}}

\noindent\textbf{Gain of the matrix accelerator.}
Figure~\ref{fig:d1}~(a) compares the core of \sys with the same core without a frontier.
On 16 threads, the gain ranges from 1.43$\times$ on the full product of the news articles to 6.88$\times$ on the gene matrix, the densest relation.
It is larger on one thread, 1.68--8.92$\times$, since the panel setup and the heavy-light path scale less well than the sparse path.
In \duckdb, the gain is up to 3.3$\times$, since the engine's scan and the consumer add the same time to both runs.
However, the accelerator matters most where \spalm, whose dense kernels suit dense blocks, is strongest.
It raises the gain of \sys over \spalm from 2.2--2.3$\times$ to 3.4--4.3$\times$ on the blocks of the most frequent news words and ratings, and from 1.7$\times$ to 5.8$\times$ on the gene matrix.
Figure~\ref{fig:d1}~(b) shows that \sys scales by 8.0$\times$ to 12.0$\times$ from one to 16 threads.

\begin{figure}[t]
\centering
\subfloat[AMX against the vector units]{\includegraphics[width=0.48\linewidth]{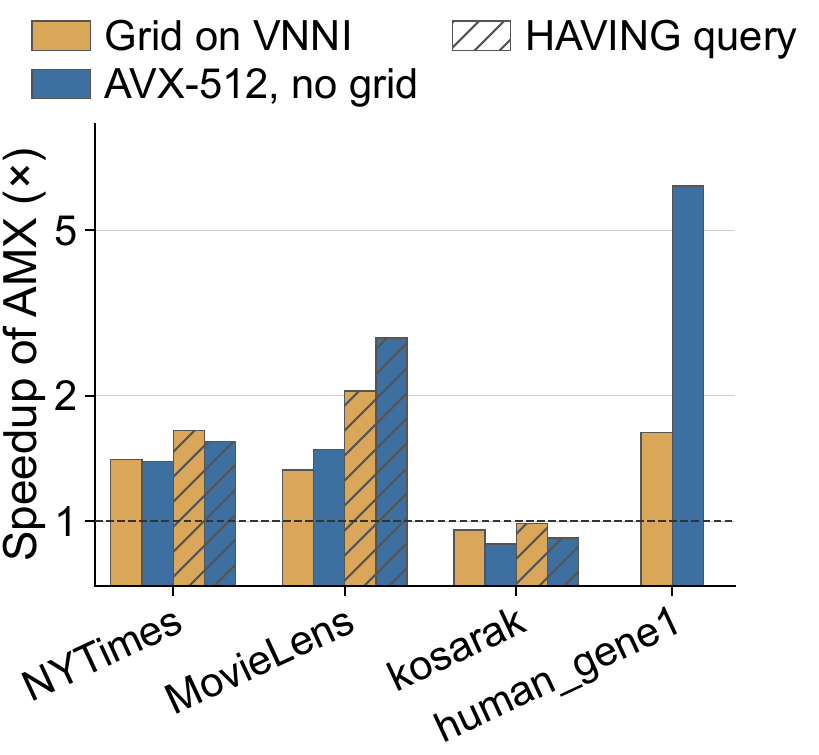}}\hfill
\subfloat[Cost model on unseen relations]{\includegraphics[width=0.48\linewidth]{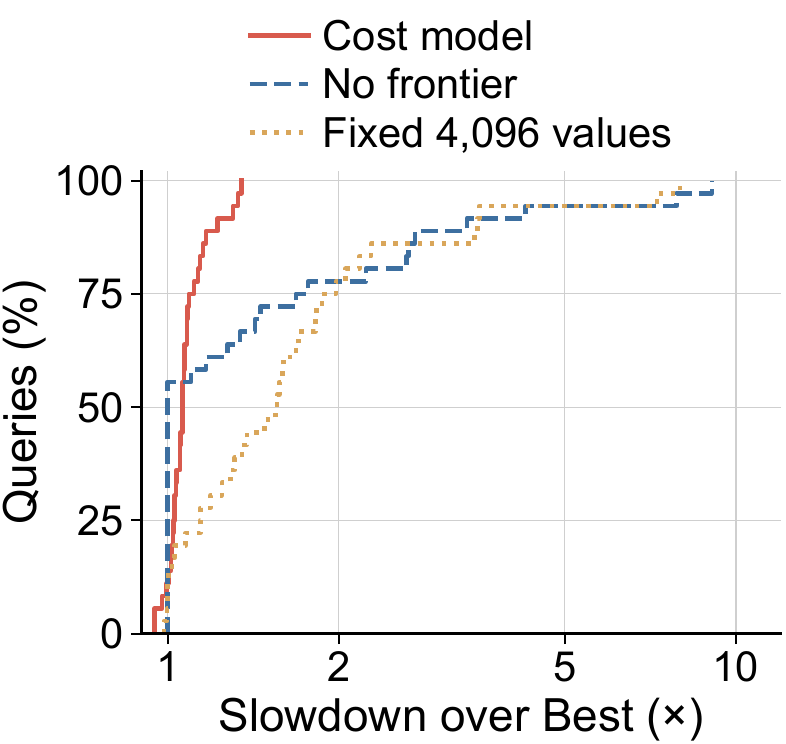}}
\caption{AMX outperforms the vector units wherever a frontier is chosen, and the cost model stays near the best frontier on relations it was not fitted to. \textit{Hatched bars are \texttt{HAVING} queries, and (b) gives the share of 36 queries within each slowdown.}\label{fig:accmodel}}
\end{figure}

\noindent\textbf{Matrix accelerator against vector units.}
Figure~\ref{fig:accmodel}~(a) compares \sys on AMX with the same grid on AVX-512 VNNI and with an AVX-512 build without a grid, which an AVX2 build matches within 6\%.
On the relations with a frontier, AMX is faster than both by 1.33$\times$ to 6.38$\times$.
This is because a tile product on VNNI costs 5.3--5.5$\times$ more, which raises the break-even density by the same factor.
As a result, the grid on VNNI gains nothing on the news articles but still 3.9$\times$ on the gene matrix, whose dense core is large.
On kosarak, the AVX-512 build is faster than AMX by 10--14\%, since AMX pays for the model's pass and then finds no frontier.

\begin{figure}[t]
\centering
\includegraphics[width=\linewidth]{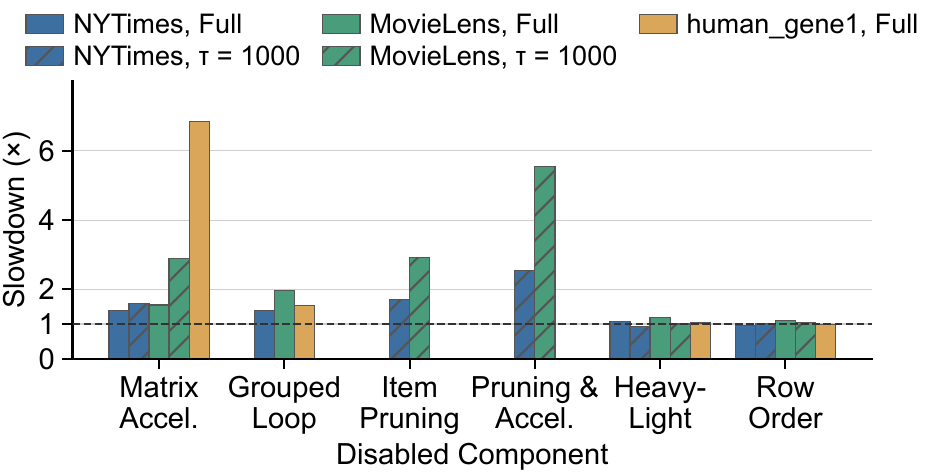}
\caption{Ablation study of \sys. The matrix accelerator, the grouped pair loop, and item pruning matter the most. \textit{Each bar is the slowdown with a mechanism disabled, against the full design at the dashed line.}\label{fig:d3}}
\end{figure}

\noindent\textbf{Ablation of each mechanism.}
Figure~\ref{fig:d3} disables one mechanism at a time, and the cost model picks the frontier of each run.
The grouped pair loop gains 1.40$\times$ to 1.96$\times$, and without it, the grid loses to running without a frontier on the full products of the news articles and the ratings.
Item pruning gains up to 2.93$\times$ on the pruned queries, and pruning and the accelerator together up to 5.53$\times$.
In contrast, the heavy-light path and the row order change the time by at most 1.19$\times$.

\begin{figure}[t]
\centering
\includegraphics[scale=0.5]{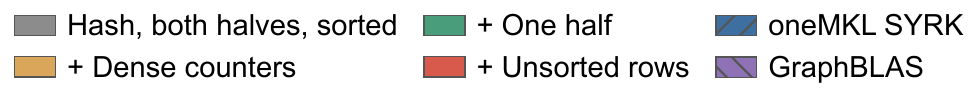}\\
\includegraphics[width=\linewidth]{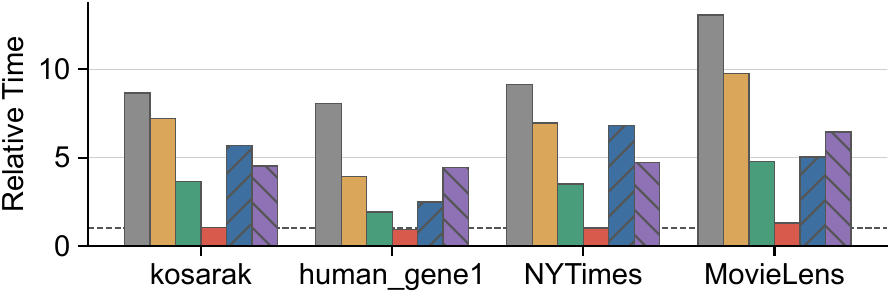}
\caption{A row-wise product with one design choice added at a time, where unsorted output rows and one half of the product gain the most. \textit{Times are relative to the sparse path of \sys.}\label{fig:ladder}}
\end{figure}

\begin{figure}[t]
\centering
\subfloat[Frontier size]{\includegraphics[width=0.48\linewidth]{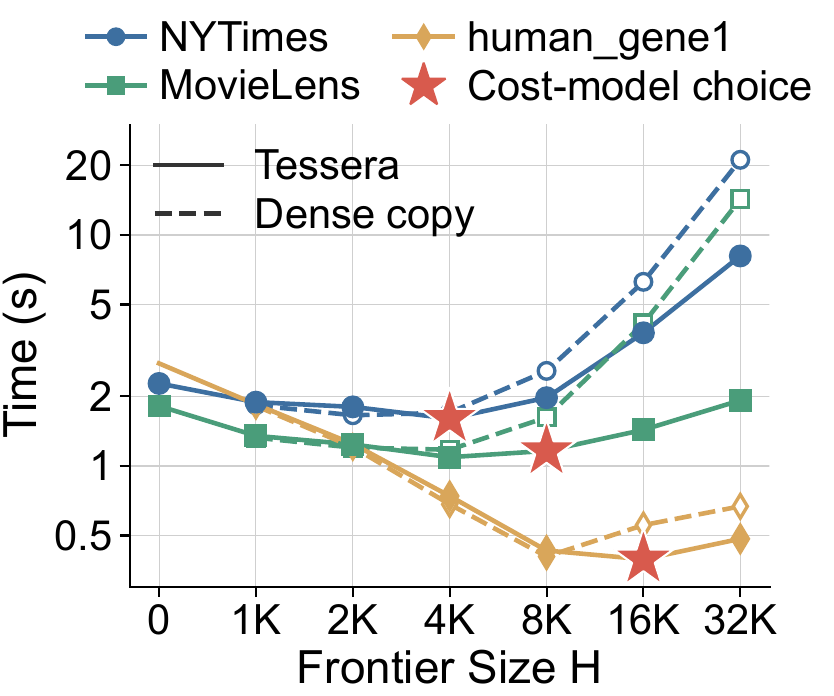}}\hfill
\subfloat[Output memory limit]{\includegraphics[width=0.48\linewidth]{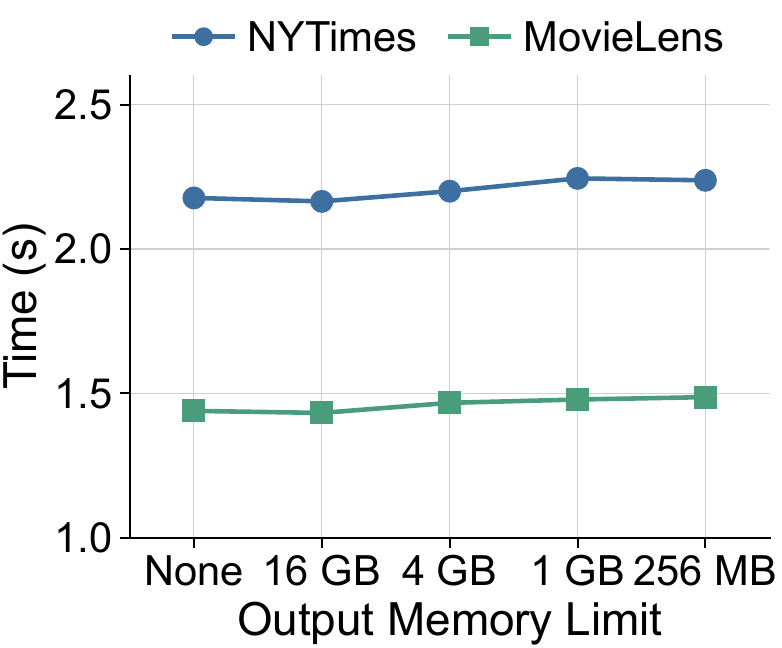}}
\caption{The cost model picks a frontier near the best fixed one, and a small output memory limit slows \sys only slightly. \textit{In (a), stars are the cost model's choices, and dashed lines a dense copy of the frontier.}\label{fig:s1}}
\end{figure}

\noindent\textbf{Breakdown of the sparse path's gain.}
Figure~\ref{fig:ladder} starts from a row-wise product with a hash accumulator that computes both halves of the result and assembles a sorted sparse matrix, and adds one design choice at a time.
Dense counters gain up to 2.1$\times$, computing one half about 2$\times$, and unsorted rows in per-thread chunks up to 3.7$\times$, which comes within 7\% of the sparse path of \sys on three relations.
This is because a hash accumulator probes a table for every product, and a sorted matrix must sort and assemble its cells before any row reaches the engine.
Both libraries fall between the first and the third step, and the sparse path outperforms them by 2.5$\times$ to 6.8$\times$.
Therefore, the gap to the libraries comes mainly from computing one half and writing rows instead of a matrix, which a relational result allows.

\noindent\textbf{Need for a per-query frontier.}
Figure~\ref{fig:s1}~(a) fixes the frontier of the full products at sizes from 0 to 32,768.
A wrong size slows the product down by up to 7.0$\times$, when it is too small, as on the gene matrix, or too large, as on the news articles.
There, the rare words give almost empty tiles that cost as much as full ones.
The best size moves from 4,096 to 16,384 values across the three products.
The cost model picks the best size on the news articles and the gene matrix and is 6.5\% slower than the best on the ratings.

\noindent\textbf{Cost model on unseen relations.}
The constants of the cost model come from four of the relations above.
Figure~\ref{fig:accmodel}~(b) runs the model on 18 other relations, including PubMed, the graphs, and six synthetic Zipf relations, with the full product and a \texttt{HAVING} query at 0.3\% support on each.
On 34 of the 36 queries, the product at the model's frontier is within 10\% of the best forced frontier.
With the model's own time, it falls behind the best by at most 1.35$\times$.
In contrast, the best fixed choice, no frontier, falls behind by up to 9.1$\times$.

\noindent\textbf{Comparison with a dense copy of the frontier.}
Figure~\ref{fig:s1}~(a) also shows, as dashed lines, a simpler design that copies the frontier into a dense matrix with one row per key and multiplies it with a vendor library's 8-bit product (oneDNN on AMX).
At its own best frontier, half the size that the cost model picks for the grid, this design comes within 5\% of the grid on the three full products.
Beyond that frontier, however, the grid outperforms it by up to 7.4$\times$.
This is because the copy has a row for every key and multiplies tiles of zeros, while the grid builds only the tiles in which rows hold values.
As a result, an oversized frontier costs the grid little, and the cost model needs no exact tuning.

\noindent\textbf{Impact of the output memory limit.}
Figure~\ref{fig:s1}~(b) limits the memory for the output of the full products in the core of \sys.
At 256~MB, \sys writes nearly all of the 571 and 883 million cells to files and is only 2.8\% and 3.3\% slower than without a limit.
This is because each thread appends whole chunks to its own file, and the reader maps the files back instead of copying them.
Hence, a result larger than the engine's memory costs little more than one that fits while its files stay in the page cache.

\begin{figure}[t]
\centering
\subfloat[PubMed, pruned queries]{\includegraphics[width=0.48\linewidth]{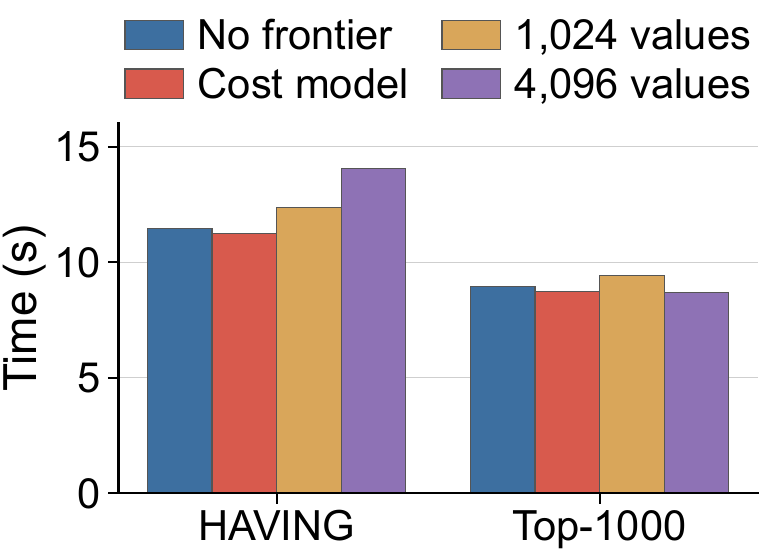}}\hfill
\subfloat[Weighted sums, over \spalm]{\includegraphics[width=0.48\linewidth]{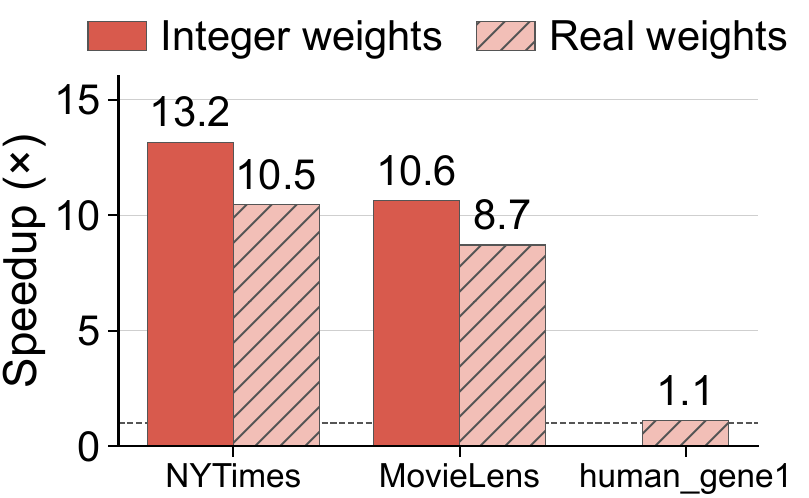}}
\caption{Where \sys does not help, a frontier without a dense block and real weights on the gene matrix.\label{fig:limits}}
\end{figure}

\noindent\textbf{Cases with little gain.}
Without a dense block, as on kosarak, PubMed, and the graphs, the cost model keeps every value on the sparse path at a cost of at most 6.2\%.
Figure~\ref{fig:limits}~(a) shows that on PubMed, frontiers of 1,024 and 4,096 values slow the \texttt{HAVING} query down by 1.08$\times$ and 1.23$\times$, respectively, since its frequent block lies below the break-even density, and the cost model chooses none.
Figure~\ref{fig:limits}~(b) shows that with real weights, which run on the floating-point sparse path, \sys outperforms \spalm by 8.7$\times$ to 10.5$\times$ on the news articles and the ratings, against 10.6$\times$ to 13.2$\times$ with integer weights.
On the gene matrix, however, \sys is only 1.1$\times$ faster, since the grid no longer takes its dense block.

%% file: sections/relatedwork.tex
\section{Related Work \label{sec:related}}

\noindent\textbf{Join-project and in-database linear algebra.}
Amossen and Pagh~\cite{amossenpagh}, Hu~\cite{humatmul}, and Deep et al.~\cite{deepjoinproject} split a join-project into heavy and light values, join the light ones, and run the heavy ones as a dense matrix product.
Deng et al.~\cite{dengoverlap} and DIM3~\cite{dim3} split sets by size or columns by density in the same way.
LevelHeaded~\cite{levelheaded} and \spalm~\cite{spalm} choose dense or sparse kernels for linear algebra inside a database.
These designs speed up the multiplication of the frequent values but leave the rare values and the result on a slow path.
\sys instead treats the rare values and the result as the main work and gives the accelerator only the dense core.

\noindent\textbf{Sparse matrix products.}
Sparse libraries~\cite{graphblas,graphblas2} compute $X^\top X$ row by row with Gustavson's algorithm~\cite{gustavson} and a sparse~\cite{gilbertspa} or hash accumulator, and some leave the rows of the product unsorted~\cite{nagasakaspgemm}.
The SYRK of oneMKL~\cite{mklsyrk} and SpSYRK~\cite{spsyrk} compute one triangle of a symmetric product.
These libraries compute a matrix outside the database, while \sys computes a relational result inside the engine and skips unneeded work.

\noindent\textbf{Pair mining and similarity joins.}
Frequent itemset miners~\cite{apriori,borgelteclat,borgeltfpgrowth,lcm} and their SQL forms~\cite{sarawagiarm} find the itemsets whose support exceeds a threshold, and similarity joins prune candidates with prefix and length filters~\cite{allpairs,ppjoin,l2ap,fiermulticore,ssjbench}.
These algorithms each serve one kind of query outside the database, while \sys brings the same pruning to SQL queries inside the engine.

\noindent\textbf{Matrix accelerators on sparse data.}
GPU kernels pack sparse rows or blocks into tensor-core tiles~\cite{tcgnn,dtcspmm,magicube,tsparse,amgt}.
CPU kernels run sparse-dense products on AMX~\cite{sparx,sparamx} and on POWER10, where DDB~\cite{ddb} sends the denser blocks of each matrix to the matrix unit.
SparseZipper~\cite{sparsezipper} proposes new matrix instructions for sparse-sparse products.
These works speed up one sparse kernel, some with new hardware, while \sys runs whole queries in an engine on existing CPUs and gives the accelerator only their dense part.

%% file: sections/conclusion.tex
\section{Conclusion \label{sec:conclusion}}

This paper presented \sys, a join-aggregate operator for database engines on skewed sparse relations.
On such relations, most of the time goes to the rare values and the output rather than to the multiplication, which limits what a faster multiplier can gain.
Therefore, \sys made these parts cheap first and gave a CPU matrix accelerator only the dense core of frequent values, which a cost model sized for each query.
In \duckdb, it outperformed the fastest existing system by 2.8$\times$ to 18.8$\times$, and in four engines, it was up to three orders of magnitude faster than their own plans.
We believe that the lesson of \sys extends beyond join-aggregates.
Matrix accelerators help a database engine most when it keeps its sparse, relational execution and hands them only the dense sub-problems that skew creates, such as the heavy hitters of joins and group-bys, rather than recasting queries as dense linear algebra.

%% file: sections/ack.tex
\section*{Acknowledgement}
This research was sponsored by the National Research Foundation of Korea (NRF) grants funded by the Ministry of Science and ICT (No. RS-2025-00522990). Gyuyeong Kim is the corresponding author.